\documentclass[twocolumn,secnumarabic,amssymb, nobibnotes, aps, prd]{revtex4-1}
\pdfoutput=1
\usepackage{amsmath}
\usepackage{nicefrac}
\usepackage{graphicx}
\usepackage{wrapfig}
\usepackage{setspace}
\usepackage{subcaption}
\usepackage{url}
\usepackage{csquotes}
\usepackage{amssymb}
\usepackage{float}
\usepackage{titlesec}
\usepackage{indentfirst}
\usepackage{mathrsfs}

\titlespacing*{\section}{0pt}{1.1\baselineskip}{\baselineskip}

\begin{document}

\title{Resonant Configurations in Scalar Field Theories: Can (Some) Oscillons Live Forever?}

\author{Marcelo Gleiser}
\email[Author's Email: ]{mgleiser@dartmouth.edu}
\affiliation{Department of Physics and Astronomy\\ Dartmouth College, Hanover, NH 03755, USA}

\author{Max Krackow}
\email[Author's Email: ]{Max.E.Krackow.GR@dartmouth.edu}
\affiliation{Department of Physics and Astronomy\\ Dartmouth College, Hanover, NH 03755, USA}

\date{\today}

\begin{abstract}
We investigate the longevity of oscillons numerically, paying particular attention to radially-symmetric  oscillons that have been conjectured to have an infinitely-long lifetime. In two spatial dimensions, oscillons have not been seen to decay. In three spatial dimensions, specific initial Gaussian configurations seem to lead to oscillons with spikes in lifetime that have been conjectured to be infinite. We study such ``resonant'' oscillons in two and three spatial dimensions, applying two tests to study their longevity: parametric resonance and virialization. Without offering a formal proof, our numerical results, within their precision, offer support for the conjecture that, in both dimensions, resonant oscillons may be infinitely long-lived.
\end{abstract}

\maketitle

\section{Introduction}
Oscillons were first discovered by Bogolyubskii and Makhan'kov in 1976 \cite{Bogolyu}.  Upon inserting a tanh profile into the three dimensional (3d) spherically-symmetric Klein Gordon (KG) equation of motion with a double well potential, the authors found that the localized configuration shed away over half its energy in a short amount of time and then settled into a metastable state.  During this quasi-steady stage, the field amplitude at the bubble core was found to gradually decrease, as energy slowly radiated away.  The name \enquote{pulson} was given to these structures since it was believed that the energy was released due to pulsations in the size of the bubble.  Two decades later, Gleiser rediscovered these pulsons \cite{Gleiser94}, showing that they emerged also from a Gaussian initial condition.  He henceforth referred to them as \enquote{oscillons} since it was shown that the radial pulsations of the bubble during the pseudo-stable regime were in fact small, and that the high frequency oscillations in the field at the bubble's core were a more notable characteristic \cite{Gleiser94}. Gleiser also showed that, irrespective of the initial profile radius (Gaussian or tanh), oscillons had the same slowly decreasing plateau energy during their lifetimes, thus suggesting that they were attractors in field configuration space. This result also holds for asymmetric double well potentials, albeit with plateau energies that depend on the asymmetry in the potential.

In \cite{Copeland}, Copeland, Gleiser and Muller (henceforth CGM) obtained a lower bound on an oscillon's initial radius, $r_0$.  Additionally, they found that once the amplitude of oscillations of the core of an oscillon falls out of the nonlinear regime, it will quickly decay, and that oscillons live longer if they are better virialized.  For a symmetric double-well potential (SDWP), the authors even caught a glimpse of the resonant structure in oscillon lifetime: certain values of $r_0$ resulted in oscillons with enhanced lifetimes.

Interested in investigating oscillon lifetimes in more detail, Honda and Choptuik \cite{Honda} (henceforth HC) developed a method using a coordinate transformation that allowed them to probe the behavior of very long-lived resonant oscillons, which we will discuss in detail in this paper.  These authors discovered a remarkable resonant pattern for the lifetime of oscillons emerging from different initial radii of Gaussian profiles, and found a scaling law for the lifetime of oscillons near each resonance reminiscent of critical behavior in phase transitions. Inspired by these results, the authors conjectured that such resonant oscillons had infinitely-long lifetimes. More recently, a similar result was found upon the incorporation of gravity, but with modulations superimposed on top of the scaling law \cite{IkedaYooCardoso}.

The longevity of oscillons has also been studied in an expanding de Sitter universe.  For different values of the Hubble constant, lifetime enhancement was seen as a consequence of parametric resonance \cite{GleiserGrahamStama1}.  For hybrid inflation models, the formation of oscillons was prompted as the inflaton field oscillated about its minimum \cite{GleiserGrahamStama2}. The authors of Refs. \cite{Amin1,Amin2,Amin3} examined the formation of oscillons for single-field inflation, and showed that they could contribute substantially to the energy density of the universe for different originating potentials. More recent work has examined the formation and longevity of oscillons for inflaton potentials of the form $|\phi|^{2n}$, near $|\phi|=0$ and flatter beyond a certain value, and their role in the post-inflationary transition from quasi de Sitter to radiation  \cite{Amin4}.

In \cite{GleiserSorn}, Gleiser and Sornborger showed that oscillons in 2d remain stable for at least $10^7$ time units, suggesting that these configurations may be infinitely long-lived.  Salmi and Hindmarsh confirmed these results and further examined 2d oscillons with several potentials, without seeing them decay \cite{SalmiHindmarsh}. Further work by G. Fodor, P. Forg\'acs, Z. Horv\'ath, and \'A. Luk\'acs \cite{Fodor1} has elucidated the relationship between quasibreathers and oscillons, while work by G. Fodor, P. Forg\'acs, Z. Horv\'ath, and M. Mezei offered an analytical approach to study the radiation rate for small-amplitude oscillons in two and three dimensions for sine-Gordon and $\phi^6$ models in 2d \cite{Fodor2}. However, the physical mechanism(s) behind the extended longevity of $\phi^4$ resonant oscillons as studied by HC remains an open question, and is the focus of this work.

Apart from an approximate analytical argument in Ref. \cite{GleiserSicilia}, there hasn't been a  consistent approach to address the possibility of {\it infinitely} long-lived oscillons. Short of a mathematical proof, what can one do to provide evidence that it is possible for certain oscillons to be infinitely long-lived, at least classically? To answer this question is the main goal of the present manuscript. In the next section, we introduce the needed mathematical framework and the monotonically increasing boosted (MIB) variables used to study very long-lived oscillons with high numerical precision. In Section III, we describe the numerical methods used. In section IV, we examine the longevity of 3d resonant oscillons using parametric resonance and virialization arguments. In section V, we examine 2d oscillons, again applying parametric resonance and virialization arguments, presenting also a toy model based on a damped, periodically-driven harmonic oscillator and phase space portraits. We conclude in section VI, reviewing our key results and suggesting future work.

\section{Oscillons: Mathematical Framework}
In this section we develop the mathematical theory behind our numerical implementation.  We follow an outline similar to HC \cite{Honda}, using metric signature (-,+,+,+) and working in spherical coordinates.
\par
The action describing the theory for a self-interacting, real scalar field in $D = d + 1$ spacetime dimensions is
\begin{equation}
S[\phi] = \int{d^Dx\sqrt{\left|g\right|}\left(-\frac{1}{2}g^{\mu\nu}\partial_{\mu}\phi\partial_{\nu}\phi - V(\phi)\right)}.
\end{equation}
We consider the following quartic potential,
\begin{equation}
V(\phi) = \frac{\lambda}{4}\phi^2\left(\phi - \phi_0\right)^2.
\end{equation}
We introduce the following dimensionless field and coordinates,
\begin{align}
\begin{split}
\phi &= \alpha\bar{\phi},\\
t &= \xi\bar{t},\\
r &= \xi\bar{r}.
\end{split}
\end{align}
If we define $v = \frac{\phi_0}{\alpha}$, $\xi^{-1} = \sqrt{\lambda} \alpha$, and drop the bars, the action becomes,
\begin{equation}
S[\phi] = \lambda^{\frac{1 - d}{2}}\phi_0^{3-d}v^{d-3}\int{d^Dx\sqrt{\left|g\right|}\left(-\frac{1}{2}g^{\mu\nu}\partial_{\mu}\phi\partial_{\nu}\phi - V(\phi)\right)}.
\end{equation}
The physical units of time and distance are then $\xi = v/\sqrt{\lambda}\phi_0$.  Choosing $v = \sqrt{2}$, our dimensionless potential can be written as
\begin{equation}
V(\phi) = \frac{1}{2}\phi^2 - \frac{1}{\sqrt{2}}\phi^3 + \frac{1}{4}\phi^4,
\label{potential}
\end{equation}
with degenerate minima at $\phi = 0$ and $\phi = \sqrt{2}$.
\par
We now present a coordinate transformation to monotonically increasing boosted (MIB) coordinates.  Our choice to transform to MIB is further motivated in section III.  We take
\begin{align}
\begin{split}
\tilde{t} &= t,\\
\tilde{r} &= r + f(r)t,\\
\tilde{\Omega} &= \Omega.
\end{split}
\label{mib}
\end{align}
Here, $f(r)$ is a smooth, monotonically increasing function interpolating between 0 and 1 at a cutoff radius, denoted $r_c$:  $f(r) \rightarrow 0$ when $r \ll r_c$ and $f(r) \rightarrow 1$ for $r \gg r_c$.  The transformation approximates spherical coordinates near the origin and light-cone coordinates as we approach spatial infinity.  This has the effect of nearly freezing and blue shifting both incoming and outgoing radiation in the region around $r_c$.  By including a dissipation term to quench high frequency modes, we can limit the amount of radiation that interferes with our numerical scheme, which is also further discussed in section III.
\par
The Euler-Lagrange equation is
\begin{equation}
\frac{1}{\sqrt{\left|g\right|}}\partial_{\mu}\sqrt{\left|g\right|}g^{\mu\nu}\partial_{\nu}\phi = \frac{\partial V}{\partial \phi}.
\end{equation}
Using the definitions
\begin{align}
a(t,r) &= 1 + f'(r)t, \\
\beta(t,r) &= \frac{f(r)}{1 + f'(r)t},
\end{align}
we have,
\begin{equation}
\begin{split}
& \frac{1}{\sqrt{\left|g\right|}}\partial_{\mu}\sqrt{\left|g\right|}g^{\mu\nu}\partial_{\nu}\phi = \frac{1}{a\tilde{r}^{d-1}}\left(\partial_t\left[a\tilde{r}^{d-1}\left(-\partial_t\phi + \beta\partial_r\phi\right)\right]\right) \\
& + \frac{1}{a\tilde{r}^{d-1}}\left(\partial_r\left[a\tilde{r}^{d-1}\left(\beta\partial_t\phi + \frac{1 - a^2\beta^2}{a^2}\partial_r\phi\right)\right]\right).
\end{split}
\end{equation}
If we define two auxiliary fields, $\Phi$ and $\Pi$ as
\begin{align}
\Phi &= \partial_r\phi, \label{auxiliaryfields1}\\
\Pi &= a\left(\partial_t\phi - \beta\partial_r\phi\right),
\label{auxiliaryfields2}
\end{align}
then we find three coupled equations of motion, one for $\phi$, another for $\Phi$, and the last for $\Pi$.  Inserting (\ref{auxiliaryfields1}) into (\ref{auxiliaryfields2}) and rearranging, we arrive at the equation of motion for $\phi$ in terms of the auxiliary fields,
\begin{equation}
\partial_t\phi = a^{-1}\Pi + \beta\Phi.
\label{eom1}
\end{equation}
The two equations for the auxiliary fields are,
\begin{align}
\begin{split}
\partial_t\Phi &= \partial_r\left[a^{-1}\Pi + \beta\Phi\right],\\
\partial_t\Pi &= \frac{1}{\tilde{r}^{d-1}}\partial_r\left[\tilde{r}^{d-1}\left(\beta\Pi + a^{-1}\Phi\right)\right] - \left(d - 1\right)\frac{f}{\tilde{r}}\Pi - a\frac{dV}{d\phi}.
\end{split}
\label{eom2}
\end{align}
In terms of the Lagrangian density, $\mathcal{L}$, the stress energy tensor is,
\begin{equation}
\begin{split}
T^{\mu\nu} &= -2\frac{\delta\mathcal{L}}{\delta g_{\mu\nu}} + g^{\mu\nu}\mathcal{L},\\
&= \partial^{\mu}\phi\partial^{\nu}\phi - g^{\mu\nu}\left(\frac{1}{2}g^{\alpha\beta}\partial_{\alpha}\phi\partial_{\beta}\phi + V\right).
\end{split}
\end{equation}
From this, we derive that the energy density, $\rho = T^{00}$, is
\begin{equation}
\rho = T^{00} = \frac{\Pi^2 + \Phi^2}{2a^2} + V(\phi).
\end{equation}
To get a better sense of the contributions to the total energy density, the gradient energy density term is
\begin{equation}
g^{00}\left(\frac{1}{2}g^{\alpha\beta}\partial_{\alpha}\phi\partial_{\beta}\phi \right) = \frac{\Pi^2 - \Phi^2}{2a^2},
\end{equation}
and the kinetic energy density term is
\begin{equation}
\partial^{0}\phi\partial^{0}\phi = \frac{\Pi^2}{a^2}.
\end{equation}

\section{Numerical Methods}
In this section we discuss our numerical coding scheme that approximates the analytical theory developed in section II.
\par
We start with a static one-dimensional lattice of $N = 2000$ points using a spacing of $\Delta r = 0.03$ and time step of $\Delta t = \Delta r / 2$, which satisfies the Courant condition.  We tested several other lattice sizes in the range $1500 \leq N \leq 10000$ and other spacing values in the ranges $0.04 \leq \Delta r \leq 0.006$, and, $0.02 \leq \Delta t \leq 0.003$.  Additionally, we tested our results using $\Delta t = \Delta r / 5$.  The evolution of our system was not affected by any of these changes.
\par 
Next, we evolve three fields, $\phi$ and the two auxiliary fields, $\Phi$ and $\Pi$, in space and time.  Our initial condition for the field, $\phi$, is a Gaussian profile of the form, $\phi(t = 0) = \sqrt{2}e^{-\left(r/r_0\right)^2}$.  This assumes that the core field starts in the $\phi = \sqrt{2}$ vacuum, and at spatial infinity the field approaches the $\phi = 0$ vacuum.  Spatial derivatives are computed using a finite difference method with second order accuracy.  For the time progression, we implement an iterative method outlined as follows.
\par
For a field, f, we advance temporally under the approximation
\begin{equation}
\partial_t\text{f}^{\, t} = \frac{\text{F}\left[\text{f}^{\, t+1}\right] + \text{F}\left[\text{f}^{\, t}\right]}{2},
\label{finitedifference}
\end{equation}
where $t$ is the time at the current time-step, and $\text{F}$ operates on $\text{f}^{\, t}$.  The action of F operating on f is to give the right hand side of Equations (\ref{eom1}) and (\ref{eom2}).  Thus, for example, we have
\begin{equation}
\partial_t\phi^t = \frac{\left[a^{-1}\Pi + \beta\Phi\right]^{t+1} + \left[a^{-1}\Pi + \beta\Phi\right]^{t}}{2}.
\end{equation}
To find $\text{f}^{\, t+1}$, we also approximate the left hand side of Equation (\ref{finitedifference}) as $\partial_t\text{f}^{\, t} = \left(\text{f}^{\, t+1} - \text{f}^{\, t} \right)/\Delta t$.  Inserting this into (\ref{finitedifference}), we arrive at the following second order scheme:
\begin{equation}
\text{f}^{\, t+1} = \text{f}^{\, t} + \left(\frac{\Delta t}{2}\right)\text{F}\left[\text{f}^{\, t}\right] + \left(\frac{\Delta t}{2}\right)\text{F}\left[\text{f}^{\, t+1}\right],
\end{equation}
and we make the definition
\begin{equation}
\text{f}_{(0)}^{\, t} \equiv \text{f}^{\, t} + \left(\frac{\Delta t}{2}\right)\text{F}\left[\text{f}^{\, t}\right].
\end{equation}
With this definition, our iterative method is
\begin{equation}
\begin{split}
\text{f}^{\, t+1} & = \text{f}_{(0)}^{\, t} + \left(\frac{\Delta t}{2}\right)\text{F}\left[\text{f}^{\, t+1}\right] \\
& = \text{f}_{(0)}^{\, t} + \left(\frac{\Delta t}{2}\right)\text{F}\left[\text{f}_{(0)}^{\, t} + \left(\frac{\Delta t}{2}\right)\text{F}\left[\text{f}^{\, t +1}\right]\right] = \cdots
\end{split}
\label{finitedifference2}
\end{equation}
We continue calculating $\text{f}^{\, t+1}$ iteratively through (\ref{finitedifference2}) until it does not change appreciably.  That is, we stop the iteration when the $l^2$-norm of $\text{f}_{(k)}$ and $\text{f}_{(k+1)}$ is less than some chosen tolerance.  Here, $k$ is the number of times the iteration has been performed.  Thus, we have that
\begin{equation}
\text{f}_{(k+1)}^{\, t+1} = \text{f}_{(0)}^{\, t} + \left(\frac{\Delta t}{2}\right)\text{F}\left[\text{f}_{(k)}^{\, t+1}\right].
\end{equation}
\par
Now we discuss the effect of incorporating MIB coordinates \cite{Honda}, and then motivate why we use them.  We have $\tilde{r} = r + f(r)t$, where $f(r) \rightarrow 0$ near $r = 0$, and $f(r) \rightarrow 1$ as $r \rightarrow \infty$.  Choosing $f(r) = \frac{1}{2}\text{tanh}\left(\frac{r-R}{\delta}\right) - \frac{1}{2}\text{tanh}\left(\frac{-R}{\delta}\right),$ we have that our MIB coordinates transition from regular spherical coordinates to light cone coordinates at a distance $R$ from the origin in a space of thickness $\delta$.  The transition to light cone coordinates blue shifts radiation and causes it to take a longer time to bounce back, effectively freezing the radiation in the region of thickness $\delta$.  The blue shift occurs because the wave packets stay the same size, but take up less space, and the time delay for radiation is a consequence of the coordinate lines becoming more nearly parallel with the world lines of the radiation.  We add a dissipative term such that the blue shifted radiation is quenched.  Thus, our numerical scheme takes the form
\begin{equation}
\text{f}^{\, t+1} = \text{f}_{(0)}^{\, t} + \left(\frac{\Delta t}{2}\right)\text{F}\left[\text{f}^{\, t + 1}\right] + \mu_{diss}\left[\text{f}^{\, t}\right].
\end{equation}
In the Equation for the field, $\phi$, the dissipative term, $\mu_{diss}$, takes the form $-\varepsilon\nabla^4\phi\Delta x^3$.  We use $\varepsilon = 0.2$. In $k$-space, we have $\nabla^4\phi \rightarrow k^4\phi_k$.  Thus, our dissipative term becomes significant at higher $k$ modes and will quench the blue shifted frequencies as mentioned above.
\par
Transforming to MIB coordinates allows us to explore $r_0$ parameter space with high resolution while still maintaining efficient computational time. 

\section{Lifetime of Resonant Oscillons}

\begin{figure}[H]
   \centering
   \includegraphics[width=0.5\textwidth]{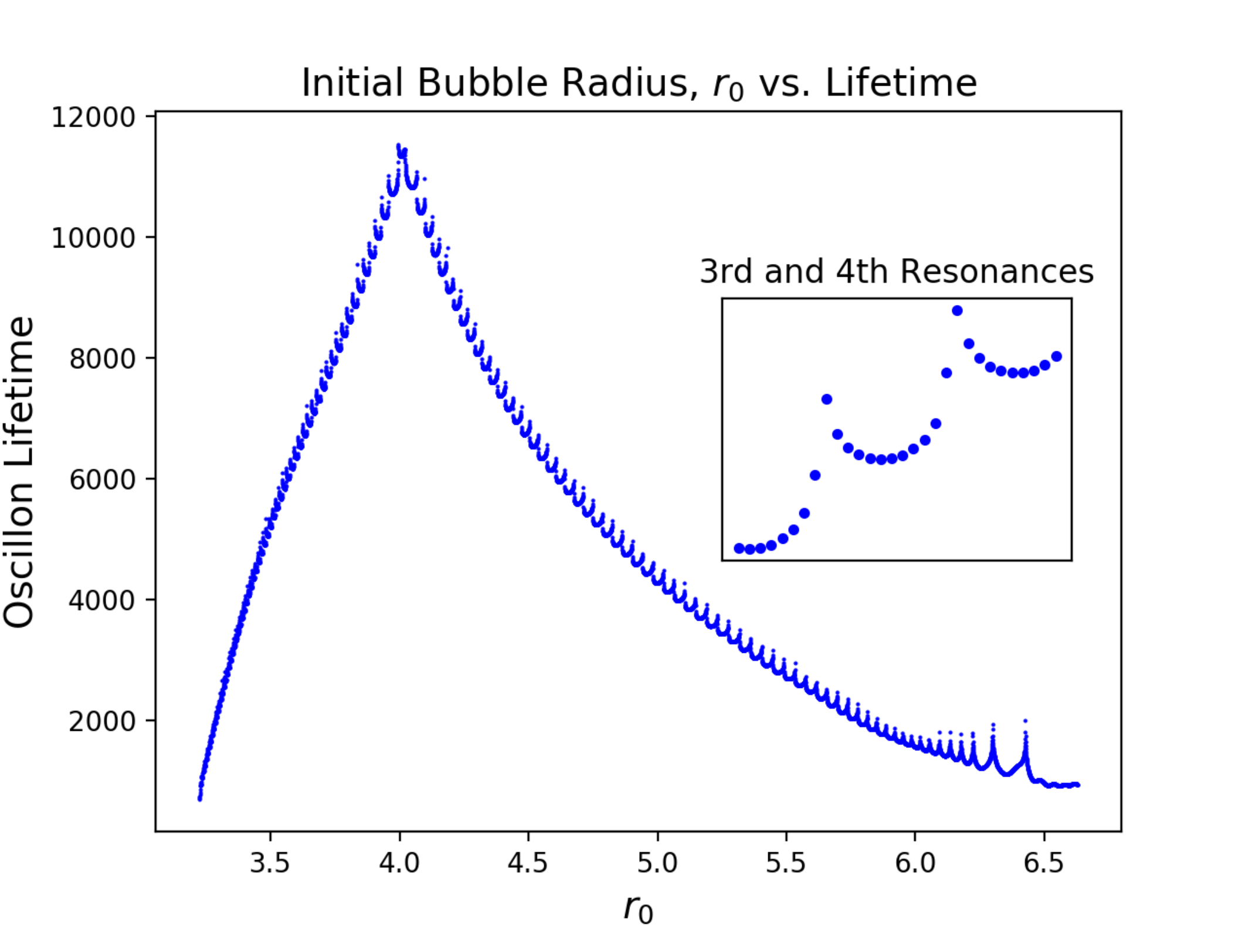}
\caption{For 3d oscillons, 127 resonances are seen along the lifetime vs. r0 plot, making the 64th resonance the highest, or peak, resonance. HC found 125 resonances in \cite{Honda} with a different choice of potential energy function. The inset shows initial radii straddling the 3rd and 4th resonances.}
\label{resmountain}
\end{figure}

Throughout section IV we will be discussing resonant oscillons and their longevity.  To get a better sense of these particular oscillons, in Figure \ref{resmountain} we plot oscillon lifetime as a function of initial bubble radius, $r_0$, with spacing $\Delta r_0=0.00047$.  The spikes in lifetime are the resonances we are referring to.  From now on we will refer to the mountain-like structure as the resonance mountain.  The inset zooms in on initial radii around the third and fourth resonances.

\begin{figure}[H]
   \centering
   \includegraphics[width=0.5\textwidth]{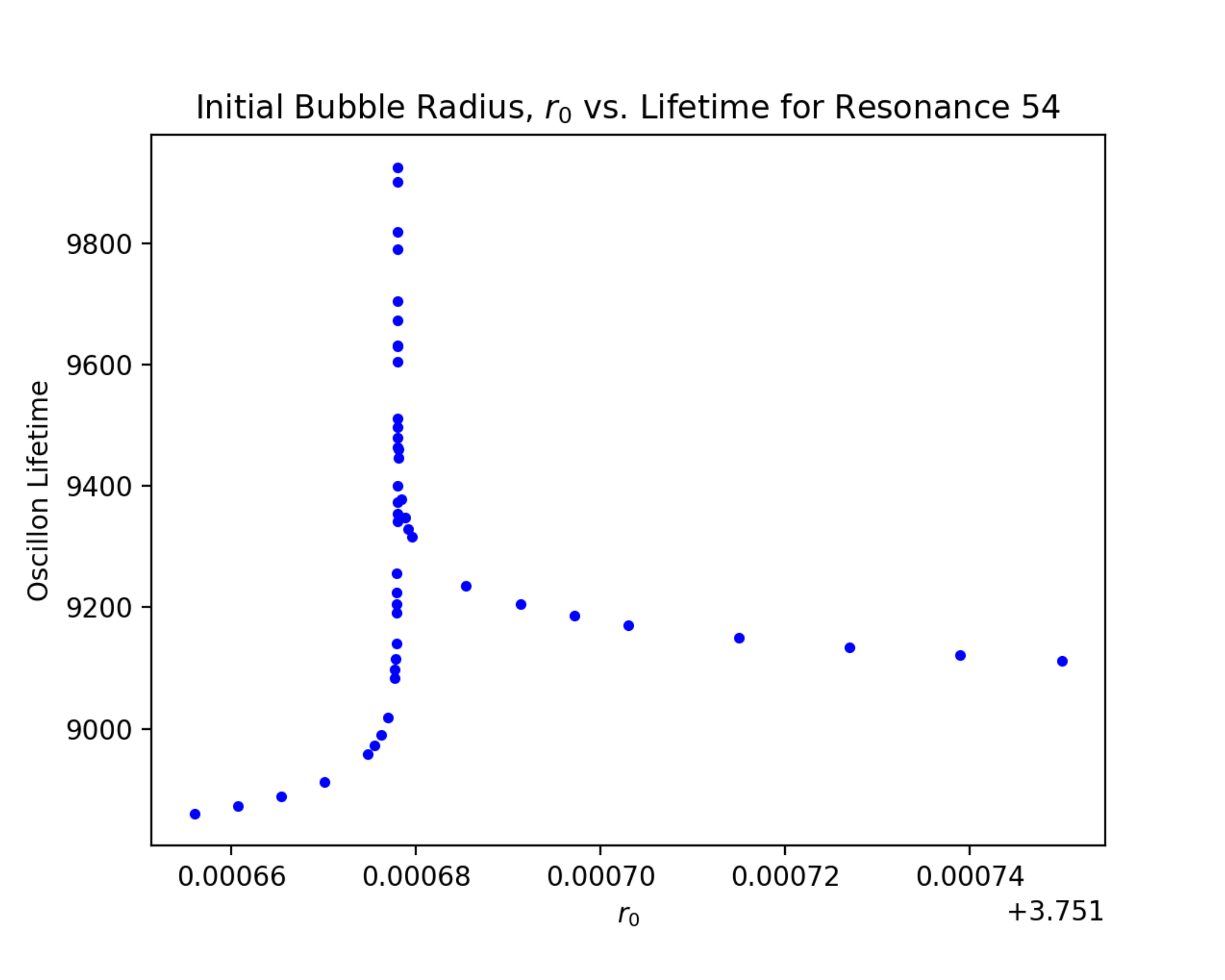}
\caption{Zoom-in on many points along resonance 54 for 3d oscillons, exploring increase in lifetime as we climb the resonance.  The resonance lies near $r_0$ = 3.75168.}
\label{res54}
\end{figure}

In Figure \ref{res54} we plot many points along the 54th resonance.  The longest-lived oscillon with a lifetime of $\tau$ = 9925 has an initial radius, $r_0$ = 3.75167794221. The numerically achievable gains are unfortunately not in the many orders of magnitude.  Compared to the oscillons with the smallest and largest inital radii in the plot, the oscillon with the maximum lifetime achieved on this resonance has a 12.0$\%$ and an 8.9$\%$ increase in lifetime respectively.  One may ask why nearby configurations do not evolve to resonant configurations as is evident in Figure \ref{res54}.  In \cite{GleiserSicilia}, the authors show that slow radiation precludes a nearby configuration from reaching the true attractor configuration in 3d. 

\subsection{Parametric Resonance Analysis}
Copeland, Gleiser, and M\"uller \cite{Copeland} found a hint of the fine structure in oscillon longevity.  The MIB framework later developed by HC \cite{Honda} allowed them to investigate the fine structure of parameter space in far greater detail.  In their search, the authors discerned 125 resonances.  They found a scaling law for lifetime of an oscillon on either side of a resonance, $T \sim \gamma_\pm\ln\left|r_0 - r_0^*\right|$, where $r_0^*$ is a resonant value.  Here, $\gamma_+$ is the scaling exponent for $r_0 > r_0^*$, and $\gamma_-$ is the scaling exponent for $r_0 < r_0^*$.  We see that as we let $r_0$ approach $r_0^*$, the scaling law implies that such oscillons are infinitely long-lived. HC then sought non-radiative solutions with an ansatz of the form
\begin{equation}
\phi\left(r,t\right) = \phi_0(r) + \sum\limits_{n=1}^\infty\phi_n(r)\cos\left(n\omega t\right).
\label{ansatz}
\end{equation}
\par
Upon truncating this ansatz, the authors used the shooting method to find an approximate solution to their equations.  They compared the individual modes from their shooting method solution with the modes obtained from Fourier decomposing the PDE solution and found a close correspondence.  Similar work in varying spatial dimensions was performed in \cite{SaffinTranberg}, where an ansatz nearly identical to that in (\ref{ansatz}), only differing by a factor of $1/\sqrt{2}$ in the zero mode, was used to construct quasi-breather solutions.  The expansion was truncated at $n=4$, and the mode profiles corresponding to particular frequencies were found numerically.
\par
Instead of performing an analysis like that in \cite{Honda} and \cite{SaffinTranberg}, we used an ansatz of the form
\begin{equation}
\phi(\textbf{x},t) = \phi_{\rm av}(t) + \delta\phi(\textbf{x},t),
\label{phifluct}
\end{equation}
as suggested originally by Gleiser and Howell \cite{GleiserHowell} and then used in other works as well \cite{GleiserGrahamStama1,GleiserGrahamStama2,Amin4}.  Here, $\phi_{\rm av}(t)$ is the volume averaged field, which is given by $\frac{3}{R^3} \int_0^R \phi\left(r,t\right) r^2 dr$.  We found that varying $R$ in the range $2r_0 \leq R \leq 5r_0$ had little effect on our results.  We insert this ansatz into the spherically symmetric Klein-Gordon equation,
\begin{equation}
\frac{\partial^2\phi}{\partial t^2} - \frac{\partial^2\phi}{\partial r^2} - \frac{2}{r}\frac{\partial\phi}{\partial r} = -\frac{dV}{d\phi},
\end{equation}
with $V(\phi)$ given in Equation (\ref{potential}). Upon taking the Fourier transform of the equation and linearizing with respect to $\delta\phi$, we are left with the relation
\begin{equation}
\delta \ddot{\phi} + \left[ k^2 + V''(\phi_{\rm av}(t))\right]\delta\phi = 0,
\label{FT}
\end{equation}
where $V''(\phi_{\rm av}(t)) = 1 - 3\sqrt{2}\phi_{\rm av} + 3\phi_{\rm av}^2$, and dots denote derivatives with respect to time.
\par 
As discussed in \cite{Honda}, dynamical properties of the field $\phi$ are approximately periodic with modulations superimposed on them.  These modulations correspond to radiation from the field and are referred to as shape modes.  Oscillons with $r_0 > r_0^*$ are referred to as supercritical since they exhibit an extra shape mode just before decaying. Those with $r_0 < r_0^*$ are called subcritical and decay without the appearance of an extra shape mode. When climbing a single resonance on the resonance mountain, we find that the only obvious difference in the behavior of $V''(\phi_{\rm av}(t))$ corresponding to different initial radii occurs at the end of the lifetime of the oscillons.  This can be seen in Figure \ref{v_double_prime} below, which displays two supercritical oscillons on the 54th resonance.  The visible difference is that one oscillon lives longer (blue curve) than the other (red curve).  Apart from this, while the two oscillons remain, their behaviors are nearly identical. For both supercritical and subcritical oscillons, there is a plateau region later in the lifetime of the oscillons where there are no shape modes. 

\begin{figure}[H]
   \centering
   \includegraphics[width=0.5\textwidth]{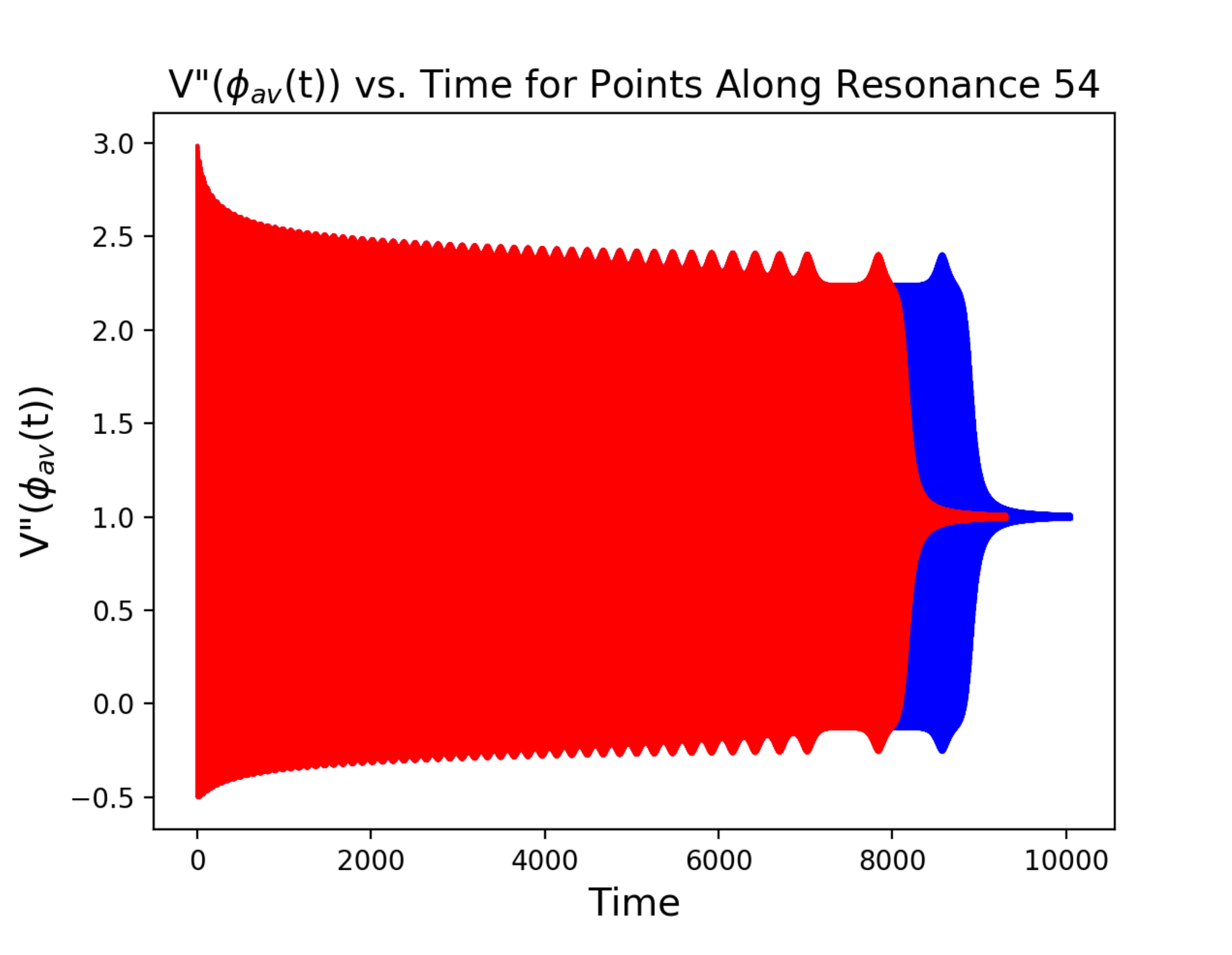}
\caption{$V''\left(\phi_{\rm av}(t)\right)$ for a low point on the 54th resonance (red) and for a point high up on the 54th resonance (blue) for 3d oscillons.  Differing behavior occurs only between the last two shape modes.}
\label{v_double_prime}
\end{figure}
\par
We can also see the similarity in behavior described above by studying the energy of the oscillons.  Although oscillons at different points along the resonance mountain radiate their energy away at different rates, oscillons within a single resonance radiate their energy away at seemingly identical rates up until they approach the end of their life.  To illustrate, we consider the energies of two oscillons also in resonance 54 with $r_0 = 3.7516972$ (green curve) and $r_0 = 3.75167794241$ (yellow curve) shown in Figure \ref{Energy_vs_Time} below.  The plateau region corresponding to the oscillon with $r_0 = 3.7516972$   lies between the black and magenta stars, and the one corresponding to the oscillon with $r_0 = 3.75167794241$ lies between the black and cyan stars.  The energy of each of these oscillons remains nearly identical until roughly $7700$ time units.  Between $7700$ and $8100$ time units, the energies begin to diverge, even if by less than $0.07 \% $ at $8100$ time units.  Just after $8100$ time units, the oscillon lower in the resonance, with $r_0 = 3.7516972$, begins to rapidly radiate away its remaining energy.  The oscillon with $r_0 = 3.75167794241$ doesn't rapidly decay until after $8700$ time units.

\begin{figure}[H]
   \centering
   \includegraphics[width=0.5\textwidth]{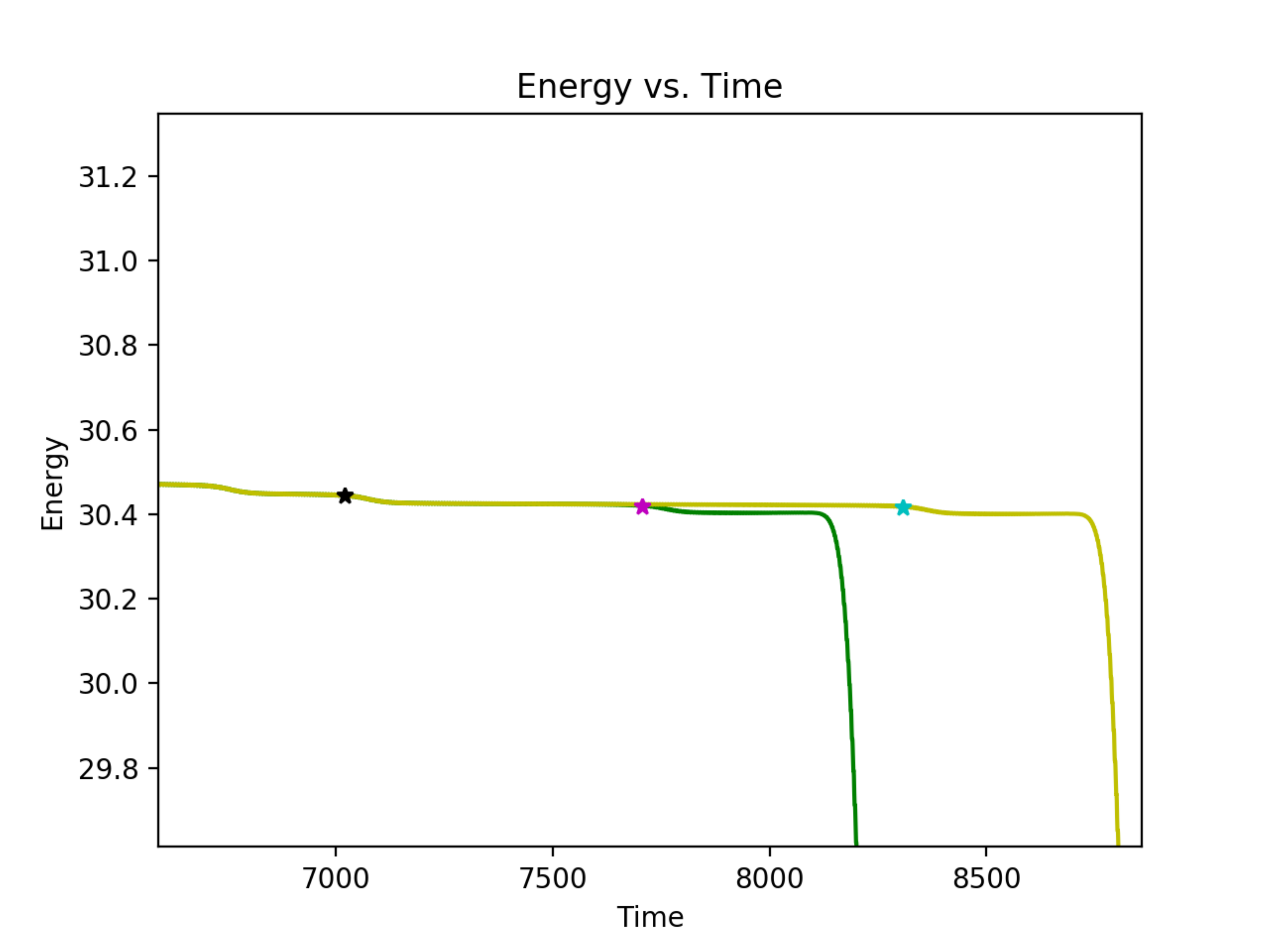}
\caption{Energy profiles of two 3d oscillons along the 54th resonance.  Slight differences in energy only become apparent near the end of the lifetimes of the two oscillons.}
\label{Energy_vs_Time}
\end{figure}

\begin{figure}[H]
	\centering
 	\begin{subfigure}[d]{0.5\textwidth}
        \includegraphics[width=\textwidth]{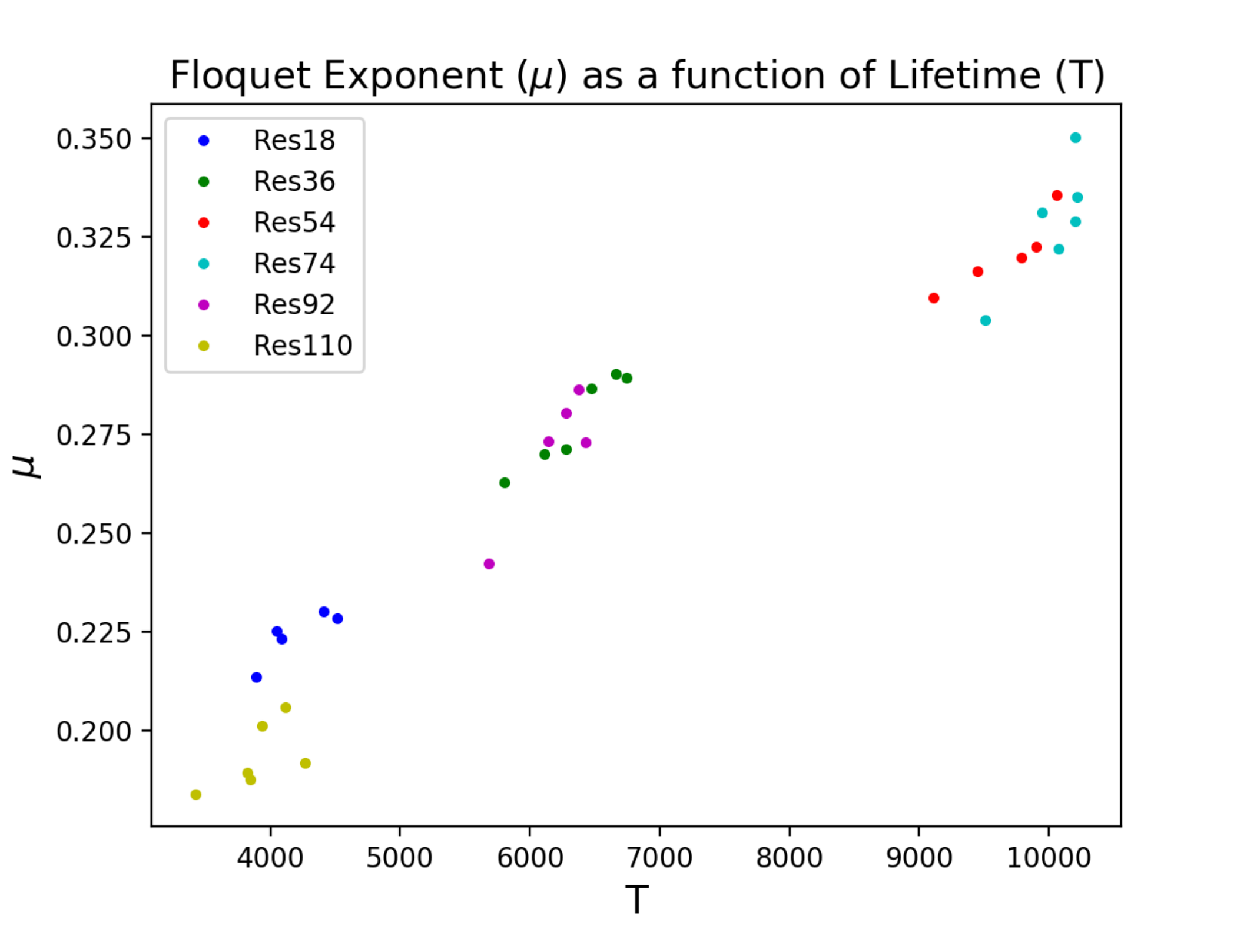}
	\caption{$\mu$ as a function of lifetime for supercritical 3d oscillons with different initial radii. Each color corresponds to a specific resonance as specified.} 
     \end{subfigure}
   \centering
    \begin{subfigure}[d]{0.5\textwidth}
       \includegraphics[width=\textwidth]{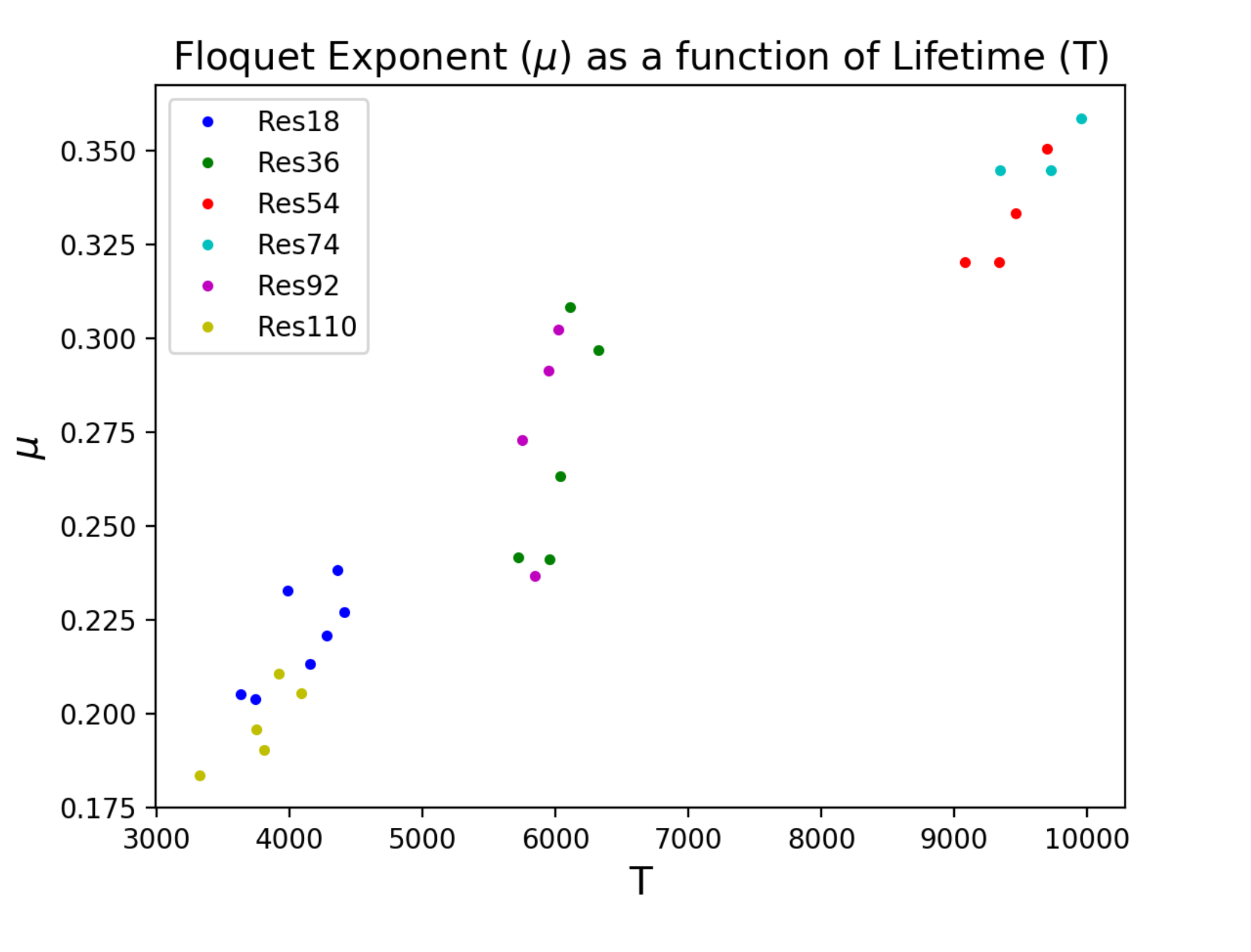}
	\caption{3d analysis: $\mu$ as a function of lifetime for subcritical 3d oscillons with different initial radii. Each color corresponds to a specific resonance as specified.}
     \end{subfigure}\caption{}
     \label{floquet}
\end{figure}

As we move higher up on a resonance, the length of the plateau region increases, and  $V''(\phi_{\rm av}(t))$ displays more nearly-periodic oscillations in this region.  Thus, for a given initial radius, we can approximate $V''(\phi_{\rm av}(t)) \approx \Phi_0 \cos(\omega t) + \textrm{C}$ for the time interval in the plateau region.
Using this approximation and making the transformation $\omega t = 2z - \pi$, we get
\begin{equation}
\delta \phi '' + \left(A_k - 2q\cos (2z)\right) \delta\phi = 0,
\label{Mathieu}
\end{equation}
where $A_k = \frac{4(k^2 + \textrm{C})}{\omega^2}$, $q = \frac{2\Phi_0}{\omega^2}$, and the primes denote derivatives with respect to $z$.  This equation is the well-known Mathieu Equation, which exhibits parametric resonance for a range of $q$ when $A_k \approx n^2, \; n = 1, \: 2, \: 3\dots$ \cite{MathieuBook}.  Thus, parametric resonance can occur when $\frac{4(k^2 + \textrm{C})}{\omega^2} \approx n^2$.  Requiring $k \geq 0$, we must have that
\begin{equation}
n^2 \geq \left(\frac{4}{\omega^2}\right)\textrm{C}.
\label{inequality}
\end{equation}
For all resonances that we examined, we found that the minimum integer value for $n$ was $n_{min} = 2$.  We next sought solutions proportional to $e^{\mu t}$, which grow exponentially for real-valued $\mu$,  the Floquet exponent.  From \cite{MathieuBook}, we can approximate $\mu$ as
\begin{equation}
\mu \approx \frac{\sqrt{\left(a_n - A_k\right)\left(A_k - b_n\right)}}{2n},
\end{equation} 
where $a_n$ and $b_n$ are functions of $q$.  We found that $n = 2$ returned real values for $\mu$, but $n > 2$ produced imaginary results.  Using $n = 2$, we saw two trends, summarized in Figure \ref{floquet}. First, as we approached the peak on the resonance mountain from either side, $\mu$ increased; second, as we climbed a {\it single} resonance, $\mu$ increased.  So, in general, the longer an oscillon lives, the greater the Floquet exponent, $\mu$: the nonlinear growth of amplitude oscillations increase the system's lifetime. This was recognized in Refs. \cite{Gleiser94, Copeland}, since large amplitude fluctuations keep the field's amplitude above the inflection point of the potential and thus within the nonlinear regime. As fluctuations weaken and the field amplitude decays below the inflection point, the oscillon enters a linear regime and rapidly decays.

The linear approximation thus suggests that the enhanced longevity exhibited by resonant oscillons is a consequence of parametric resonance.  Illustrative results are displayed in Figure \ref{floquet} for points along resonances 18, 36, 54, 74, 92, and 110.  In \ref{floquet}a we display results for supercritical oscillons and in \ref{floquet}b we show our findings for subcritical oscillons.  As we climb a single resonance, the level of numerical precision becomes increasingly demanding. For example, for resonance 54, the longest lived oscillon we probed had $r_0=3.75167794221$. Still, the growing trends offers support for  the HC conjecture, even if not a definitive proof.
We note that the general trend we found between the Floquet exponents and lifetime is not perfect: a few oscillons with shorter lifetimes do exhibit larger Floquet exponents.  This may be a consequence of using a linear approximation to study the growth of different $k$-modes, as typical of any Floquet analysis. 

In Figure \ref{floquet} we notice similarities in the values of $\mu$ and lifetime between resonances 18 and 110, resonances 36 and 92, and resonances 54 and 74.  The similarities in each pairing are due to the fact that each resonance in a pairing is the same number of resonances away from the peak resonance - resonance 64.  For example, consider the pairing of resonances 18 and 110.  We have that 110 - 64 = 64 - 18 = 46.  We shouldn't expect the values of $\mu$ and lifetime for these resonance pairings to be identical, given that the resonance mountain is not symmetric about its peak, as seen in Figure \ref{resmountain}.

\subsection{Virialization}
\begin{figure}[H]
   \centering
   \includegraphics[width=0.5\textwidth]{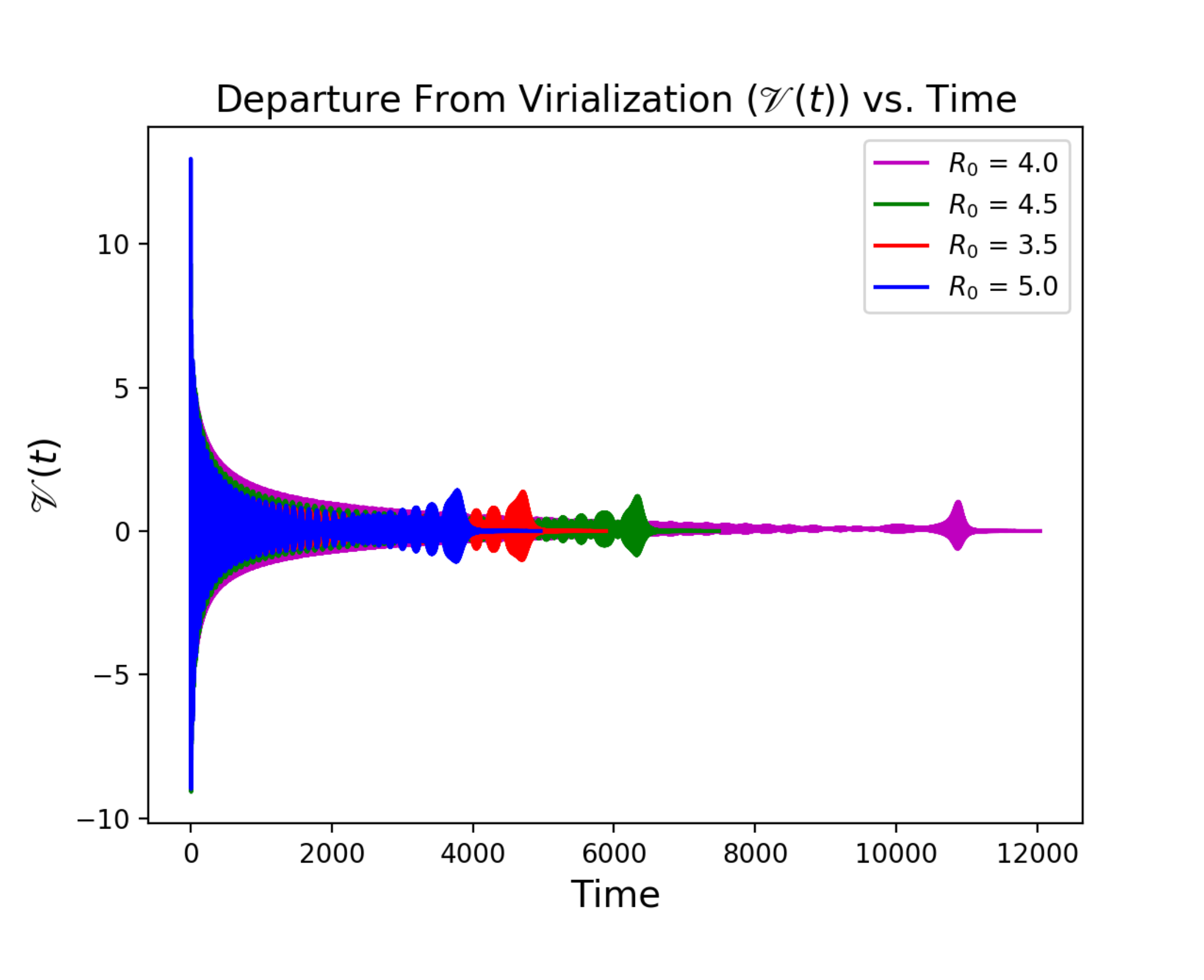}
\caption{The departure from virialization decreases for 3d oscillons higher on the mountain.  These oscillons live for longer periods of time.}
\label{virialfig}
\end{figure}

As mentioned in the introduction, oscillons that are longer-lived are also better virialized \cite{Copeland}.  This has been shown for non-resonant oscillons in flat spacetime without a transformation to MIB coordinates.  With our ability to hone in on specific resonances at high precision, we can test whether oscillons become better virialized as we climb a single resonance.  To arrive at the virial relation in \cite{Copeland}, the authors took the first moment of the equation of motion for $\phi$ and then performed a time-averaging on the equation over a single period of the core field, $\phi(0,t)$. Here, we find the frequency of oscillations of the core field by taking its Fourier transform and then computing the period, $T = 1/f$.  The departure from virialization in MIB coordinates, denoted $\mathscr{V}(t)$, is, with the potential of equation (\ref{potential}),
\begin{equation}
\begin{aligned}
&\mathscr{V}\left(t\right) = \left\langle E_k \right\rangle - \left\langle E_s \right\rangle \\
& - 2\pi \left\langle \int_0^{\tilde{R}} \tilde{r}^2 \phi^2 \left(\phi - \frac{1}{\sqrt{2}}\right)\left(\phi - \sqrt{2}\right)d\tilde{r}\right\rangle, 
\end{aligned}
\end{equation}
\noindent 
where,
\begin{equation}
\begin{aligned}
&E_k = 2\pi\int_0^{\tilde{R}}\tilde{r}^2\left(a^{-1}\Pi + \beta\Phi\right)^2d\tilde{r}, \\
&E_s = 2\pi\int_0^{\tilde{R}}\tilde{r}^2\left(\Phi\right)^2d\tilde{r}.
\end{aligned}
\end{equation}
Here, $\tilde{r}$, $\Phi$, and $\Pi$ are given by equations (\ref{mib}), (\ref{auxiliaryfields1}), and (\ref{auxiliaryfields2}), respectively.  To test our code, in Figure \ref{virialfig} we show the departure from virialization for four oscillons along the resonance mountain.  The oscillon with an initial radius $R_0 = 4.0$ is the longest lived, compared to those with initial radii $R_0 = 4.5, \: 3.5, \textrm{ and } 5.0$, respectively.  This can be confirmed by referring to Figure \ref{resmountain}.  From Figure \ref{virialfig}, we see that later in the lifetime of the oscillons, the longer lived an oscillon  the smaller $\mathscr{V}(t)$ and, thus, the better virialized it is.  These results agree with \cite{Copeland}.

When comparing the departure from virialization for different intial radii along a single resonance, it is difficult to discern by eye which oscillons  are better virialized.  To compensate for this, we define a different measure for the departure from virialization, which we call delta departure from virialization, denoted $\Delta\mathscr{V}$.  To find $\Delta\mathscr{V}$, we first define $\left|\Delta\rho(r,t)\right|$, the absolute difference in energy density between an arbitrary oscillon located on a given resonance and the longest lived oscillon whose data we have saved on that same resonance: $\left|\Delta\rho(r,t)\right| \: = \: \left|\rho_{\textrm{arbitrary}}(r,t) - \rho_{\textrm{longest lived}}(r,t)\right|$.  At every time-step in our simulation, we save the maximum value of $\left|\Delta\rho(r,t)\right|$, denoted $\textrm{max}\left(\left|\Delta\rho(t)\right|\right)$.  Next, we find the maximum of the array, $\textrm{max}\left(\left|\Delta\rho(t)\right|\right)$, which we refer to as $\textrm{max}_2\left(\left|\Delta\rho\right|\right)$.  The time index associated with $\textrm{max}_2\left(\left|\Delta\rho\right|\right)$ is our time of interest. We call it $t_{\rm max}$. In words, this is the time of maximal difference between the energy density of the arbitrary oscillon of interest on a given resonance and the energy density of the longest-lived oscillon on that same resonance whose data we have saved.  Next, we calculate the period of the core field in a narrow time region centered on $t_{\rm max}$.  We use this period to calculate the departure from virialization, $\mathscr{V}(t)$, within this same time region, and then take the difference between the maximum and minimum values of $\mathscr{V}(t)$ in this region.  This value for a given oscillon is $\Delta\mathscr{V}$.  Just as $\mathscr{V}(t)$ decreases with increasing lifetime, we expect that $\Delta\mathscr{V}$ would also decrease with increasing oscillon lifetime.

\begin{figure}[H]
	\centering
 	\begin{subfigure}[d]{0.5\textwidth}
        \includegraphics[width=\textwidth]{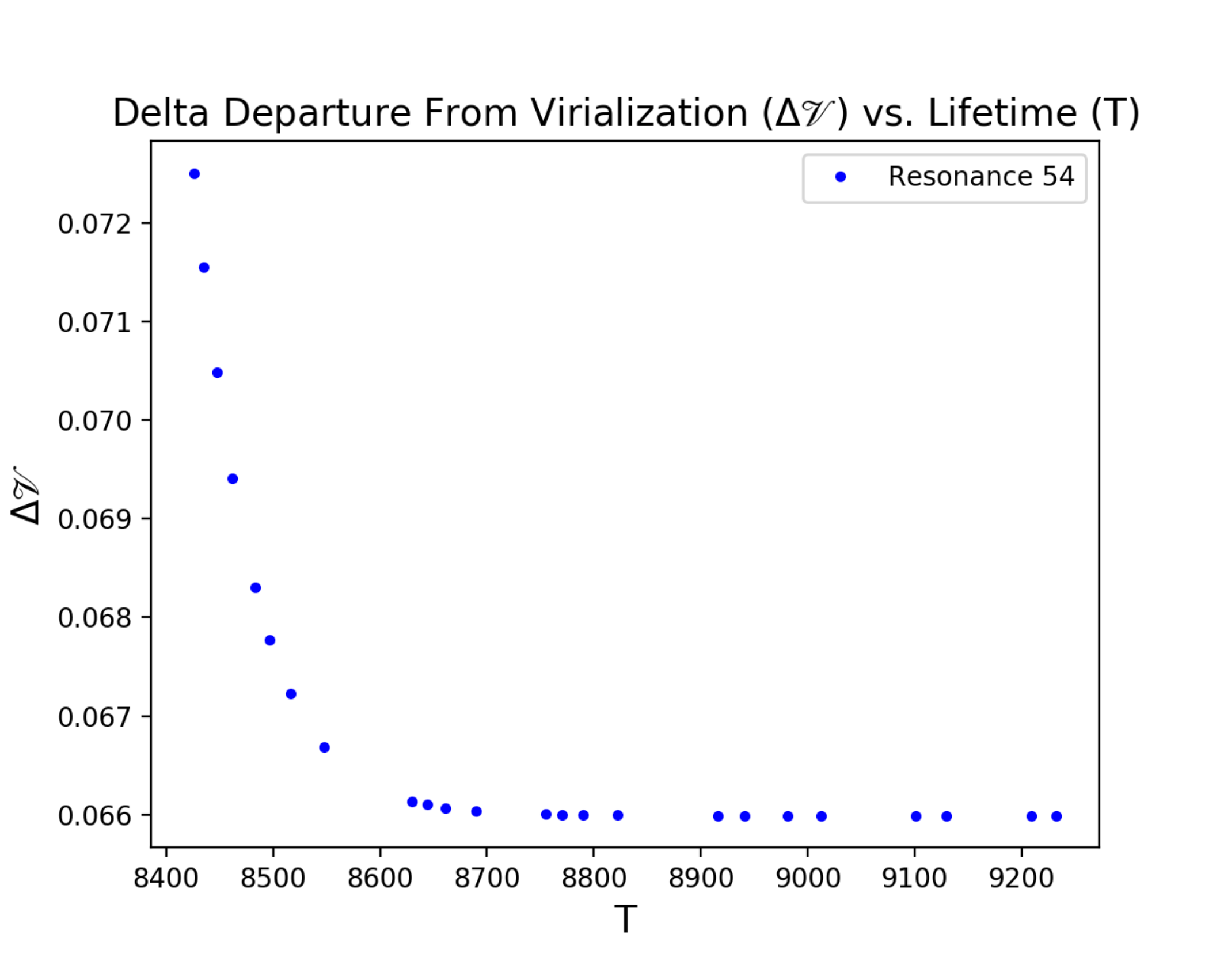}
	\caption{$\Delta\mathscr{V}$ for 24 3d oscillons along resonance 54.  $\Delta\mathscr{V}$ appears to plateau as lifetime continues to increase.} 
     \end{subfigure}
   \centering
    \begin{subfigure}[d]{0.5\textwidth}
       \includegraphics[width=\textwidth]{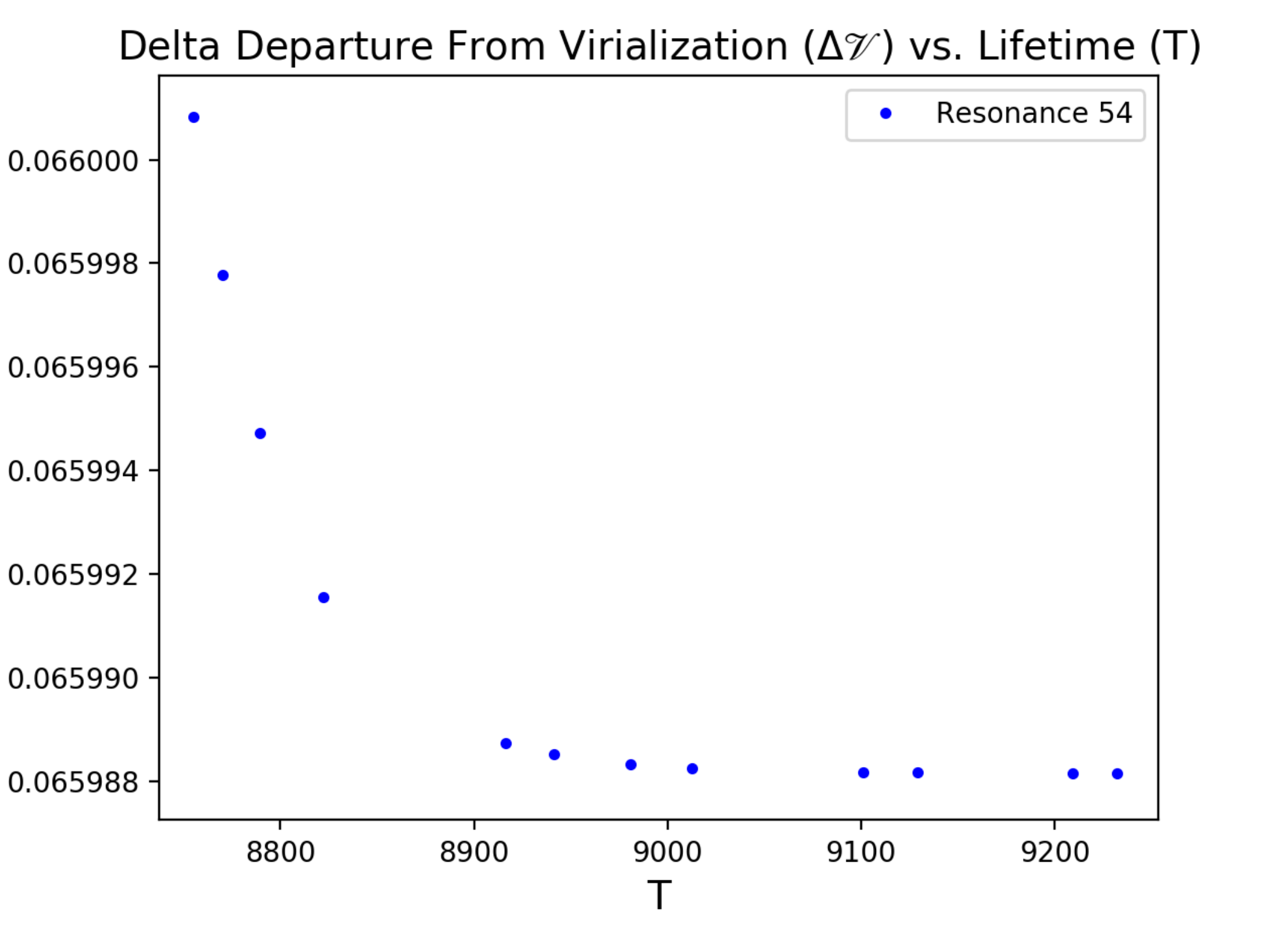}
	\caption{Zooming in on the plateau region in subplot a), we see that there is still a decrease in $\Delta\mathscr{V}$.}
     \end{subfigure}\caption{}
     \label{deltavir}
\end{figure}

Figure \ref{deltavir} below plots $\Delta\mathscr{V}$ as a function of lifetime for many supercritical oscillons along the 54th resonance.  In \ref{deltavir}a we see all data points we collected along resonance 54.  As we climb the resonance, $\Delta\mathscr{V}$ decreases monotonically until a lifetime of roughly 8,700 time units, and then it appears to plateau for the last 12 points.  In \ref{deltavir}b, we zoom in on the last 12 points of \ref{deltavir}a and find that the monotonic decrease continues followed by what again appears to be a plateau at roughly 9,000 time units.   

Successive zoom-ins show that $\Delta\mathscr{V}$ keeps decreasing as we climb the resonance, as anticipated:  longer-lived oscillons on a given resonance do have a smaller $\Delta\mathscr{V}$.  Our numerical results show a trend that supports, within the limits of our investigation, the HC conjecture: as the critical radius $r_0^*$ is approached, the related oscillon's lifetime asymptotically approaches infinity. This is reminiscent of simulating criticality in phase transitions, where a finite lattice cannot properly attain a true divergence at the critical point.

\section{Two Dimensional Oscillons}
\begin{figure}[H]
   \centering
   \includegraphics[width=0.5\textwidth]{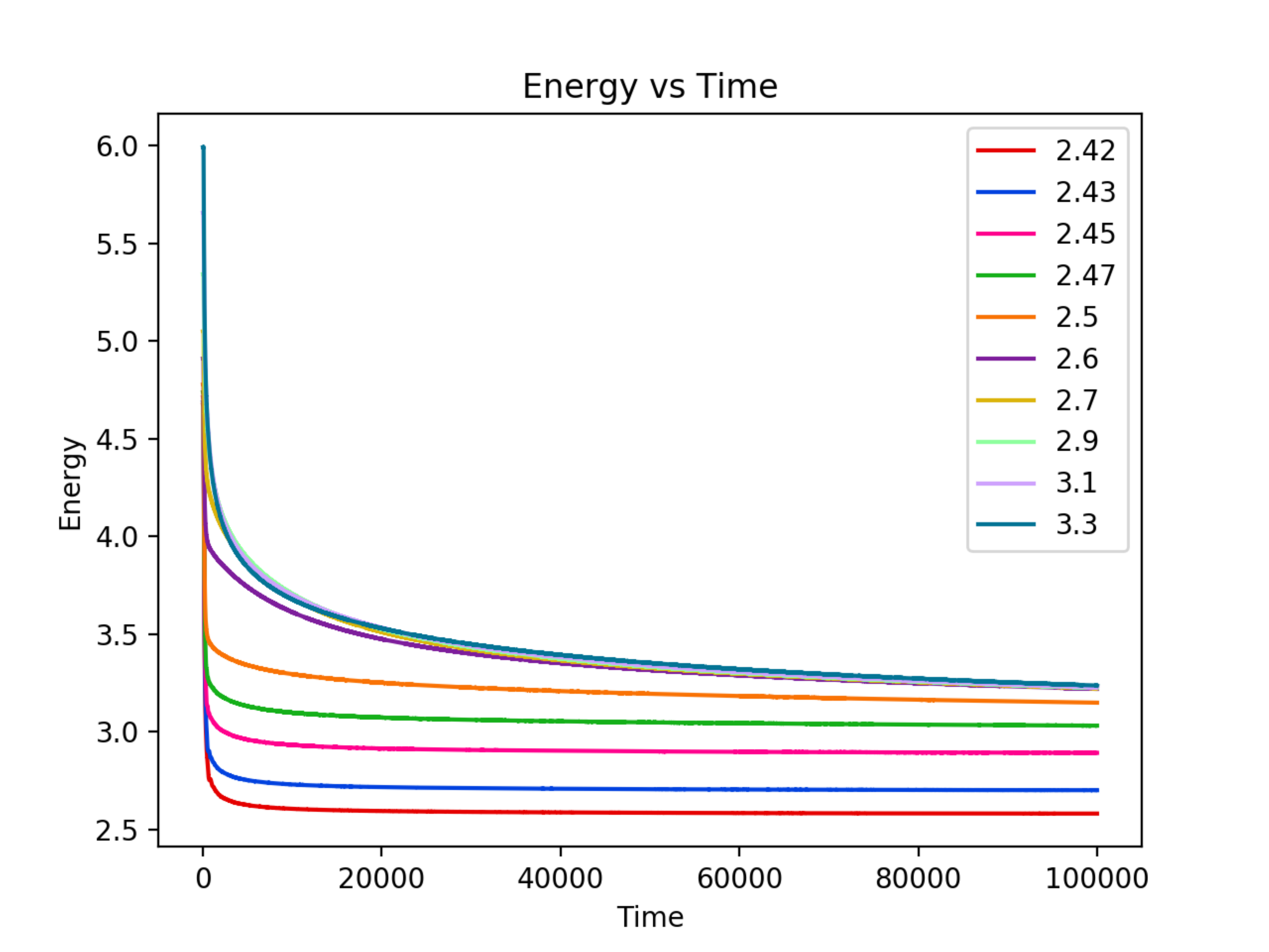}
\caption{Two-dimensional oscillon energy out to 100,000 time units.  Oscillons with different initial radii plateau at different energies.}
\label{2doscillons}
\end{figure}

For a large range of initial radii, Gleiser and Sornborger found that 2d oscillons do not decay out to at least $10^7$ time units \cite{GleiserSorn}.  Their results were confirmed by Hindmarsh and Salmi \cite{SalmiHindmarsh}. The findings suggest that 2d oscillons are stable solutions. To date, there have been no indications to the contrary, at least classically. Above, we ran several experiments in 3d to test the conjecture that resonant oscillons are infinitely long-lived.  Since 2d oscillons might never decay, it is natural to investigate their longevity with the tools we applied to 3d oscillons. 
First, we discuss results from a Floquet analysis in 2d.  Gleiser and Sornborger found that, at least for numerically feasible time-scales, not all oscillons converge to the same plateau energy in 2d, as can be seen in Figure \ref{2doscillons}, where we plot oscillon energy as a function of time for several different initial radii.

\par
The energies of oscillons with initial radii 2.42, 2.43, 2.45, 2.47, and 2.5 do not converge over 100,000 time units.  However, it appears that the energy profiles for oscillons with initial radii 2.7, 2.9, 3.1, and 3.3 do converge. A careful analysis in different time regions shows that they do, in fact, have slightly different energies over the 100,000 time units displayed. The different energies have an interesting consequence for the Floquet exponents, as we show next.
\par

\subsection{Parametric Resonance in 2d}
The equation of motion in 2d is,
\begin{equation}
\frac{\partial^2\phi}{\partial t^2} - \frac{\partial^2\phi}{\partial r^2} - \frac{1}{r}\frac{\partial\phi}{\partial r} = -\frac{dV}{d\phi},
\label{2deom}
\end{equation}
where $V\left(\phi\right)$ is given in equation (\ref{potential}), as before.  As in 3d, we write the field as in equation (\ref{phifluct}) to obtain the equation for the Fourier transform of $\delta\phi$, equation (\ref{FT}).

\begin{figure}[H]
   \centering
   \includegraphics[width=0.5\textwidth]{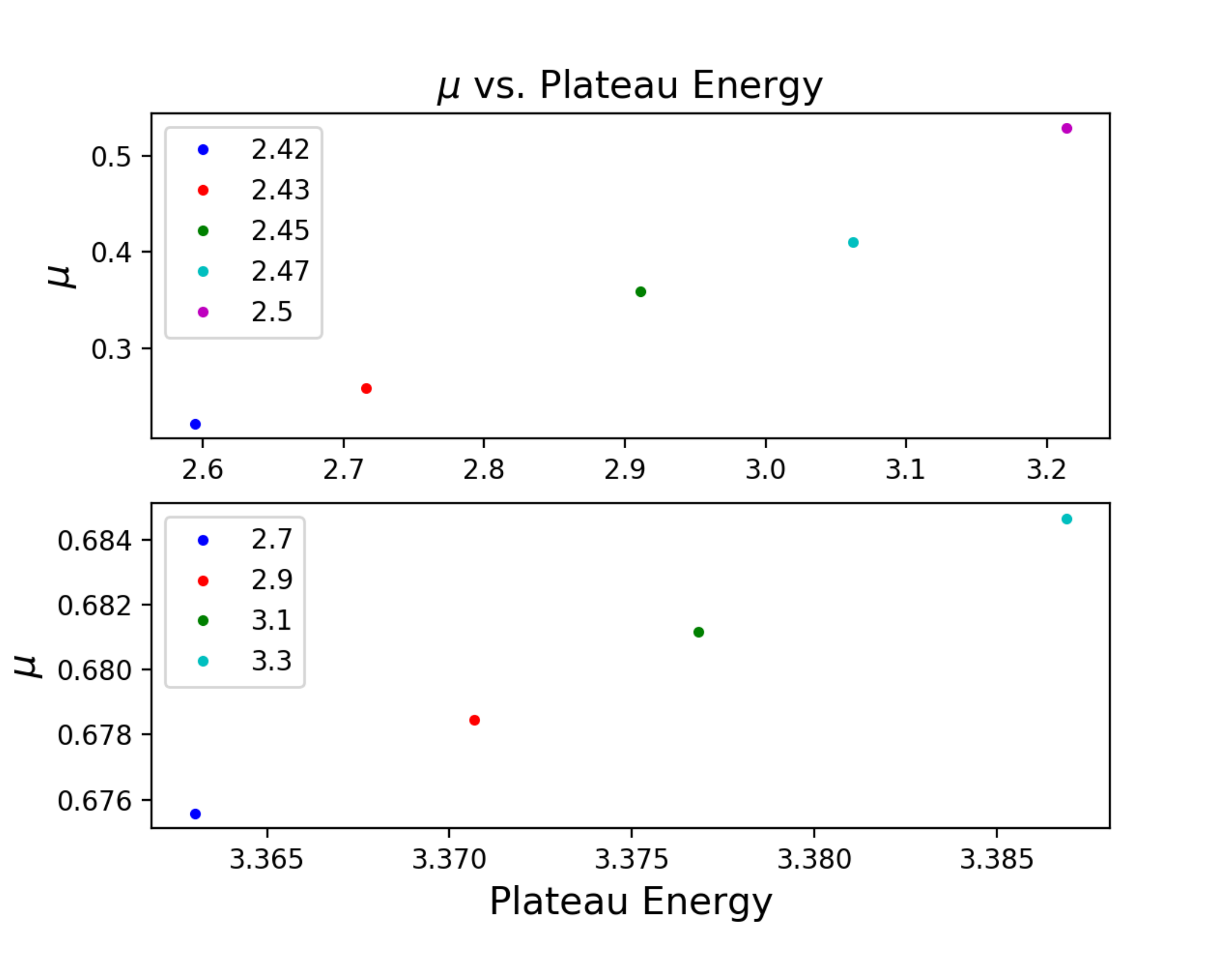}
\caption{The higher the plateau energy of an 2d oscillon, the greater its Floquet exponent, $\mu$.  The frequency that we used to calculate $\mu$ was the dominant frequency of $V''\left(\phi_{\rm av}\left(t\right)\right)$ found by taking the time Fourier transform between $39,000$ and $40,000$ time units.}
\label{floquet2d}
\end{figure}

The fact that the equations for $\delta\phi(k,t)$ are the same in 3d and 2d can be understood by examining the 3d and 2d Fourier transforms.  In 3d, the spherically symmetric Fourier transform is given by,
\begin{equation}
f\left(r\right) = \frac{1}{\left(2\pi\right)^3}4\pi\int_0^{\infty}F\left(k\right)k^2\frac{\sin{\left(kr\right)}}{kr}dk,
\end{equation}
and in 2d the azimuthally symmetric Fourier transform is,
\begin{equation}
f\left(r\right) = \frac{1}{2\pi}\int_0^{\infty}F\left(k\right)J_0\left(kr\right)kdk.
\end{equation}
Using the approximations and definitions in 2d as we used in 3d, we obtain equation (\ref{Mathieu}),
the Mathieu equation. The essential difference between 2d and 3d is encapsulated in the time-dependent frequency of $V''(\phi_{\rm av}(t))$, and thus in $A_k$ and $q$. 
We search for solutions proportional to $e^{\mu t}$, where, as before, $\mu$ is approximated as,
\begin{equation}
\mu \approx \frac{\sqrt{\left(a_n - A_k\right)\left(A_k - b_n\right)}}{2n}.
\end{equation}
As in 3d, we find that $n = 2$ satisfies the relation in equation (\ref{inequality}) and returns real values for $\mu$.  Since 2d oscillons do not decay, we cannot plot their Floquet exponents, $\mu$, as a function of lifetime.  Instead, we plot $\mu$ as a function of plateau energy, as shown in Figure \ref{floquet2d}.

The top panel of Figure \ref{floquet2d} displays the Floquet exponents for oscillons with initial radii 2.42, 2.43, 2.45, 2.47, and 2.5.  The Floquet exponents for oscillons with initial radii 2.7, 2.9, 3.1, and 3.3 are shown in the bottom panel, since they have similar plateau energies.  We see that oscillons with higher plateau energies have larger Floquet exponents.  A larger Floquet exponent means that there is greater parametric amplification, since our solutions are proportional to $e^{\mu t}$.  Comparing with Figure \ref{2doscillons}, we see that higher initial energy implies a larger initial energy loss. The larger amplitude amplification of these oscillons act to compensate for this loss, keeping the field oscillations within the nonlinear range.

\par
In 3d, the frequency that we used to calculate $\mu$ was the dominant frequency of $V''\left(\phi_{\rm av}\left(t\right)\right)$ in the plateau region, which we discussed when presenting Figure \ref{v_double_prime}.  No plateau region exists in 2d.  Instead, we can pick the dominant frequency in any time region of $V''\left(\phi_{\rm av}\left(t\right)\right)$ to calculate $\mu$.  In Figure \ref{floquet2d}, we used the dominant frequency between 39,000 and 40,000 time units.  To ensure that the relationship we found between Floquet exponent and plateau energy was not due to an arbitrary choice of time region, we performed our calculations of the Floquet exponents by looking at many different time regions.  We calculated the Floquet exponents in a narrow time region around 20,000 time units and continued calculating them in narrow time regions after every successive 5,000 time units up until we reached 100,000 time units.  We found two trends.  First, for a given time region examined, $\mu$ is greater for oscillons with higher plateau energies.  Thus, the trend found in Figure \ref{floquet2d} does not only hold between 39,000 and 40,000 time units, but holds for at least many narrow time regions between 20,000 and 100,000 time units.  Second, as we let the system evolve forward in time, $\mu$ decreases, indicating that the system becomes more stable. Figure \ref{floquetlong} below shows these two trends.  We plot only oscillons with initial radii 3.1 and 3.3 for visual clarity.  The two trends described above hold for all 2d oscillons that we studied.

\par

\begin{figure}[H]
   \centering
   \includegraphics[width=0.5\textwidth]{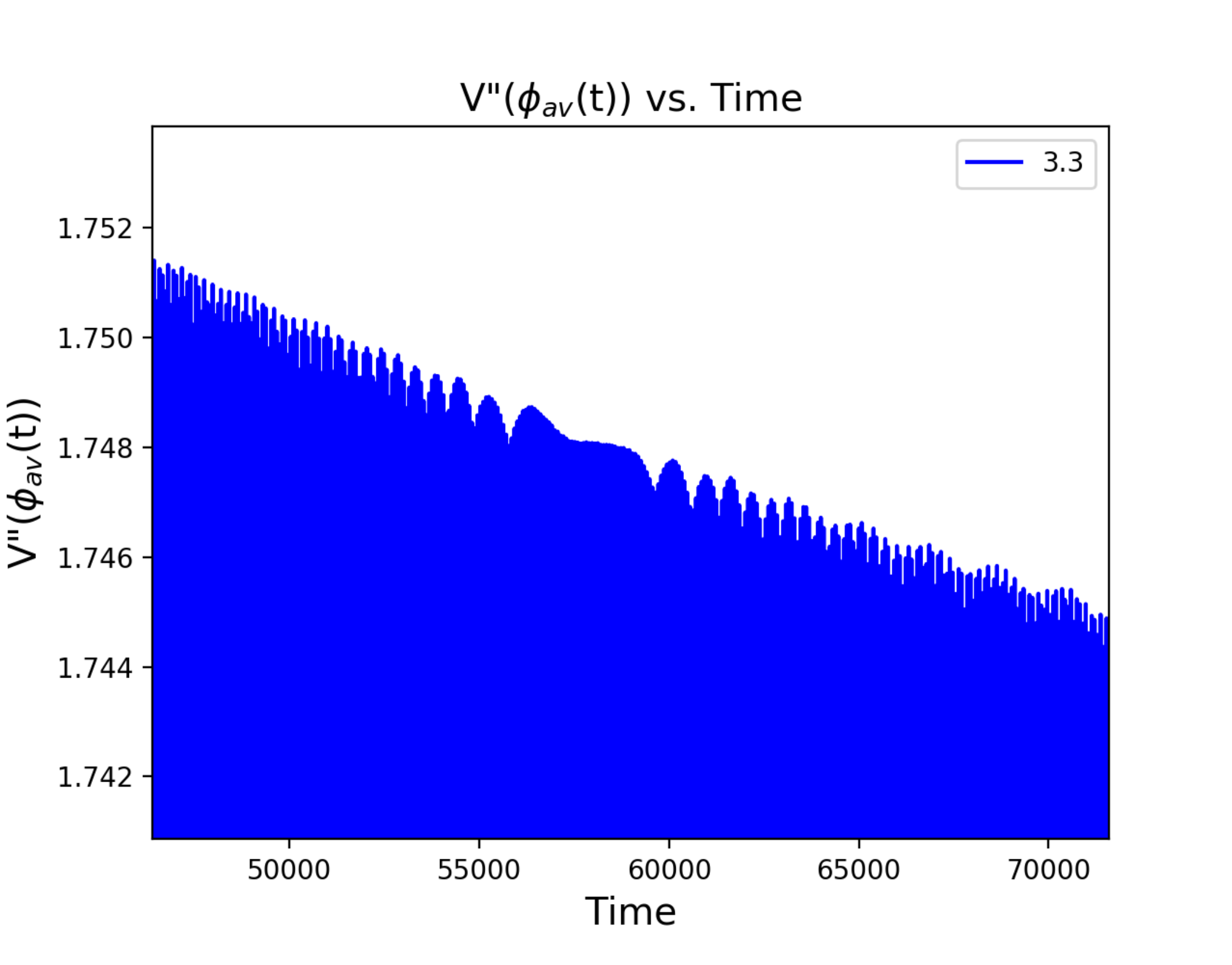}
\caption{Near the top envelope of $V''\left(\phi_{\rm av}\left(t\right)\right)$ for a 2d in the time region around 55,000 to 60,000 time units.}
\label{v''}
\end{figure}

In Figure \ref{floquetlong} we see that in addition to the Floquet exponents decreasing as a function of time, the rate at which they decrease also appears to slightly decrease over time.  For all times, the Floquet exponent of the oscillon with $r_0= 3.3$ is greater than that of the oscillon with $r_0= 3.1$.  From Figure \ref{floquet2d}, we see that the oscillon with $r_0= 3.3$ has a higher plateau energy than that with $r_0 =3.1$.  As we said before, Floquet exponents increase with increasing plateau energy.  We also note that between 55,000 and 60,000 time units there is a significant drop in the Floquet exponent for the oscillon with $r_0= 3.3$ and that between 50,000 and 55,000 time units, the Floquet exponent of the oscillon with $r_0 =3.1$ exhibits similar behavior.  To understand why this happens we looked at the behavior of $V''\left(\phi_{\rm av}\left(t\right)\right)$, since the Floquet exponents are calculated from this function.  We plot $V''\left(\phi_{\rm av}\left(t\right)\right)$ for the oscillon with  $r_0=3.3$ in Figure \ref{v''}, zooming in on the time region around 45,000 to 75,000 time units near the top envelope of the function. 

\begin{figure}[H]
   \centering
   \includegraphics[width=0.5\textwidth]{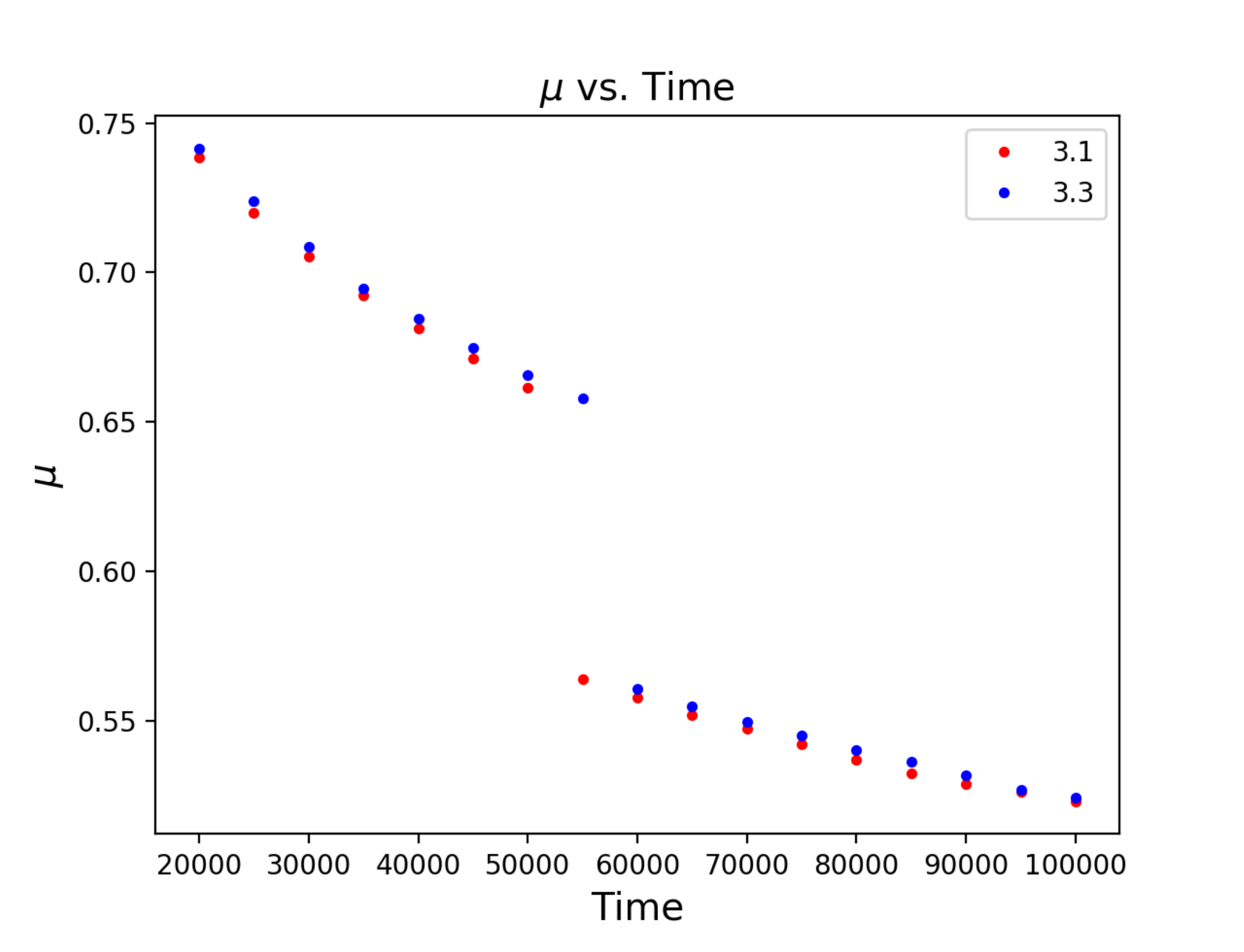}
\caption{As time evolves, a 2d oscillon's Floquet exponent decreases.  Oscillons with higher plateau energies maintain higher Floquet exponents for all times.}
\label{floquetlong}
\end{figure}

\par

Between 55,000 and 60,000 time units the fast frequency of $V''\left(\phi_{\rm av}\left(t\right)\right)$ changes abruptly. This correlates with the sudden drop in Figure \ref{floquetlong}. A similar structure is present in $V''\left(\phi_{\rm av}\left(t\right)\right)$ between 50,000 and 55,000 time units for the oscillon with $r_0= 3.1$. We have not observed any other such change of behavior in the time range between 20,000 and 100,000 time units.

\subsection{Damped, Driven Oscillator Analogy}
We wish to see if we can relate qualitatively the behavior found for the Floquet exponents $\mu$ in 2d to a familiar system -- the harmonic oscillator.  Consider a damped harmonic oscillator with time-dependent spring constant,
\begin{equation}
x'' + \gamma x' + \left[ \omega_0^2 - \frac{F_0}{mx_0} \cos{\left(\omega z\right)} \right]x = 0.
\end{equation}
Here, $\gamma$ is the damping constant of the oscillator.  If we let $x\left(z\right) = e^{-\alpha z}y\left(z\right)$ and set $\gamma = 2\alpha$, we are left with,
\begin{equation}
y'' + \left[\omega_0^2 - \alpha^2 - \frac{F_0}{mx_0}\cos{\left(\omega z\right)}\right]y = 0.
\end{equation}
Further, if we let $A_k = \omega_0^2 - \alpha^2$, $q = \frac{F_0}{2mx_0}$, $\omega = 2$, and identify $\delta \phi\left(z\right)$ with $y\left(z\right)$, we have the Mathieu Equation written in the same form as in equation (\ref{Mathieu}).
\par
Since solutions to the Mathieu Equation are of the form $e^{\mu z}P\left(z\right)$, where  $P\left(z\right)$ is a periodic function, we obtain,
\begin{equation}
x\left(z\right) = e^{\left(\mu - \gamma /2\right)z}P\left(z\right).
\end{equation}
\par
If $\mu$ is greater (smaller) than $\gamma /2$, the oscillation amplitude grows (decays) exponentially.
\par
In our 2d oscillon system, $\mu$ decreases as a function of time, and the rate at which it decreases also decreases over time as the oscillons become more stable.  In this toy model, an infinitely long-lived oscillator would start with  $\mu > \gamma /2$, with $\mu \rightarrow \gamma /2$ for large times, so that $x\left(z\right)$ becomes periodic.
\par
Since, for oscillons, $\mu$ decreases over time and $\mu$ is a function of $q$, we also have that $q$ decreases in time.  Relating this to the harmonic oscillator system, as $q$ decreases, $\frac{F_0}{mx_0}$ decreases.  This means that, over time, the oscillator's spring constant experiences smaller-amplitude fluctuations:  the oscillator behaves more and more like an oscillator with time-independent spring constant, in qualitative analogy with long-lived 2d oscillons. This corroborates the increased virialization we found in 3d and, as we shall see next, in 2d as well.
\par

\subsection{Virialization in 2d}
We start by taking the first moment of the equation of motion of our system, which is given by (\ref{2deom}).  That is, we multiply through by $2\pi r\phi$ and integrate over all space.  After integrating by parts, we get,  
\begin{equation}
2\pi\int_0^{\infty} \left(r\phi \ddot{\phi} + r\left(\phi '\right)^2 + r\phi\frac{\partial V}{\partial\phi}\right)dr = 0.
\end{equation}
Next, we perform a time-averaging and again integrate by parts,
\begin{equation}
\frac{2\pi}{T}\int_t^{t + T}\int_0^{\infty}\left(-r\dot{\phi}^2 + r\left(\phi '\right)^2 + r\phi\frac{\partial V}{\partial\phi}\right)drdt = 0.
\label{virial2d}
\end{equation}
Denoting the time-averaging by,
\begin{equation}
\left\langle \: \: \: \right\rangle \equiv \frac{1}{T}\int_t^{t + T}dt,
\end{equation}
and noting that the first term in equation (\ref{virial2d}) is minus twice the kinetic energy, and that the second term is twice the surface energy, we are left with the following relation,
\begin{equation}
\left\langle E_k \right\rangle  = \left\langle E_s \right\rangle + \pi\left\langle \int_0^{\infty}r\phi\frac{\partial V}{\partial\phi}dr\right\rangle.
\end{equation}
We define the departure from virialization in 2d as,
\begin{equation}
\mathscr{V}\left(t\right) = \left\langle E_k \right\rangle  - \left\langle E_s \right\rangle - \pi\left\langle \int_0^{\infty}r\phi\frac{\partial V}{\partial\phi}dr\right\rangle.
\end{equation}
Using the MIB transformations of equation (\ref{mib}) and the potential of Equation (\ref{potential}),
we obtain,
\begin{equation}
\begin{aligned}
& \mathscr{V}\left(t\right) = \left\langle E_k \right\rangle  - \left\langle E_s \right\rangle \\
& - \pi\left\langle \int_0^{\tilde{R}}\tilde{r} \phi^2 \left(\phi - \frac{1}{\sqrt{2}}\right)\left(\phi - \sqrt{2}\right)d\tilde{r}\right\rangle,\\
\end{aligned}
\end{equation}
where,
\begin{equation}
\begin{aligned}
& E_k = \pi\int_0^{\tilde{R}}\tilde{r}^2\left(a^{-1}\Pi + \beta\Phi\right)^2d\tilde{r}; \\
& E_s = \pi\int_0^{\tilde{R}}\tilde{r}^2\left(\Phi\right)^2d\tilde{r}.
\end{aligned}
\end{equation}

\par
Since oscillons do not decay in 2d, we ran tests to see if there was a relationship between the departure from virialization and plateau energy.  We found that the departure from virialization increases with increasing plateau energy.  To show this, we define a relative departure from virialization between an oscillon with $r_0$ and the other with $r_0=2.42$,  which has the lowest plateau energy of the oscillons we studied: $\Delta \mathscr{V} = 
\mathscr{V}_{r_0}(t) - \mathscr{V}_{r_0=2.42}(t)$.

\begin{figure}[H]
\includegraphics[width=0.5\textwidth]{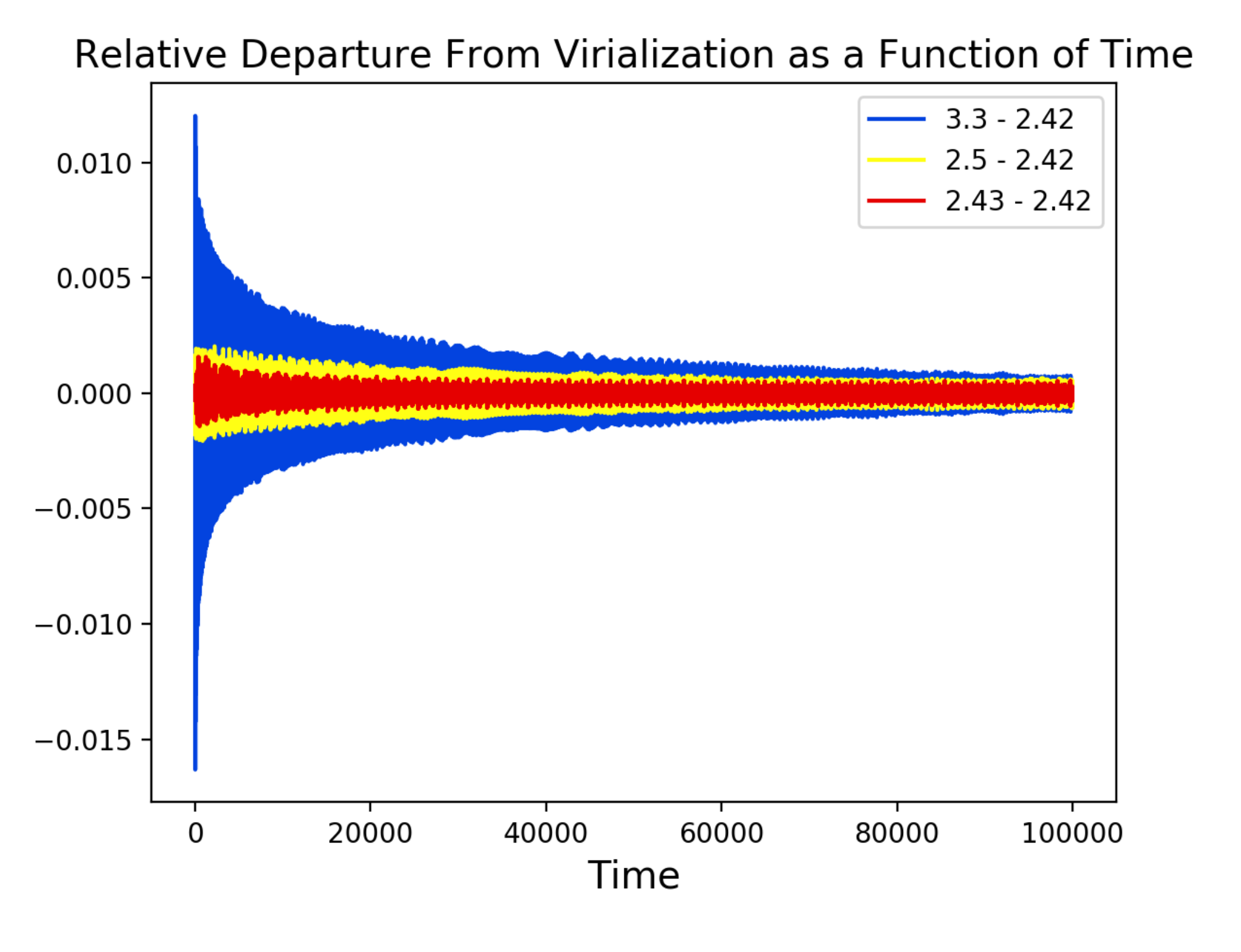}
\caption{The relative departure from virialization is plotted as a function of time for different initial radii of 2d oscillons.  The higher the plateau energy, the greater the relative departure from virialization.  Note that the relative departure from virialization decreases over time.}
\label{relvir}
\end{figure}

\par
Figure \ref{relvir} displays the relative departure from virialization as a function of time for oscillons with initial radii 2.43, 2.5, and 3.3.  We plot only three curves for visual clarity.  The increase in relative departure from virialization with higher plateau energy correlates with the increase in the parametric instability $\mu$, and with how oscillons with larger initial energies approach semi-harmonic stability asymtoptically. This suggests that the lowest energy oscillon with $r_0=2.42$ behaves as an attractor in field configuration space, as hypothesized in Ref. \cite{GleiserSicilia}.
This trend holds true for all oscillons that we studied in 2d.  We also note that the amplitude of the relative departure from virialization decreases in time, thus pointing towards greater stability as time evolves, which is again in agreement with our parametric resonance analysis results.  

\subsection{Phase Space in 2d}
We have shown two trends that 2d oscillons exhibit.  One is that oscillons with higher plateau energies are more unstable in the short term; the other is that, contrary to na\"ive expectations, 2d oscillons become more stable over time.  Plotting their phase space diagrams supports these trends.  

\begin{figure}[H]
   \centering
   \includegraphics[width=0.5\textwidth]{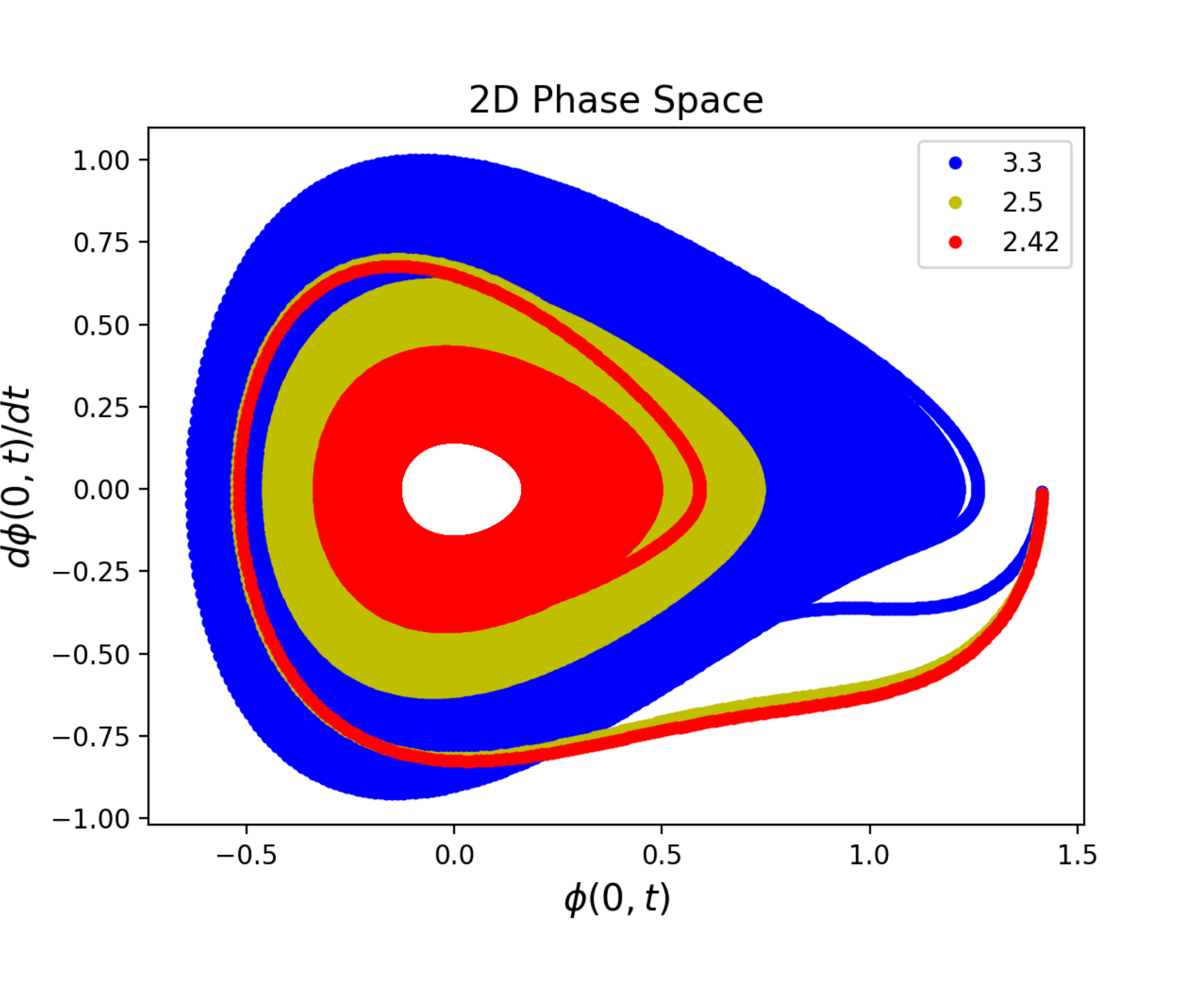}
\caption{Phase space diagram showing that 2d oscillons with higher plateau energies occupy larger regions in phase space.}
\label{phasediagram1}
\end{figure}

Figure \ref{phasediagram1} is a phase space diagram for oscillons with initial radii 2.42 (red), 2.5 (yellow), and 3.3 (blue).  As in the virialization section, we plot the phase space of only three different oscillons for visual clarity.  Each oscillon is plotted for 100,000 time units.  Comparing Figure \ref{phasediagram1} with Figure \ref{2doscillons}, we see that oscillons with higher plateau energies occupy a larger area in phase space. This trend between the plateau energy and the area of phase space holds for all 2d oscillons that we studied.

\begin{figure}[H]
   \centering
   \includegraphics[width=0.5\textwidth]{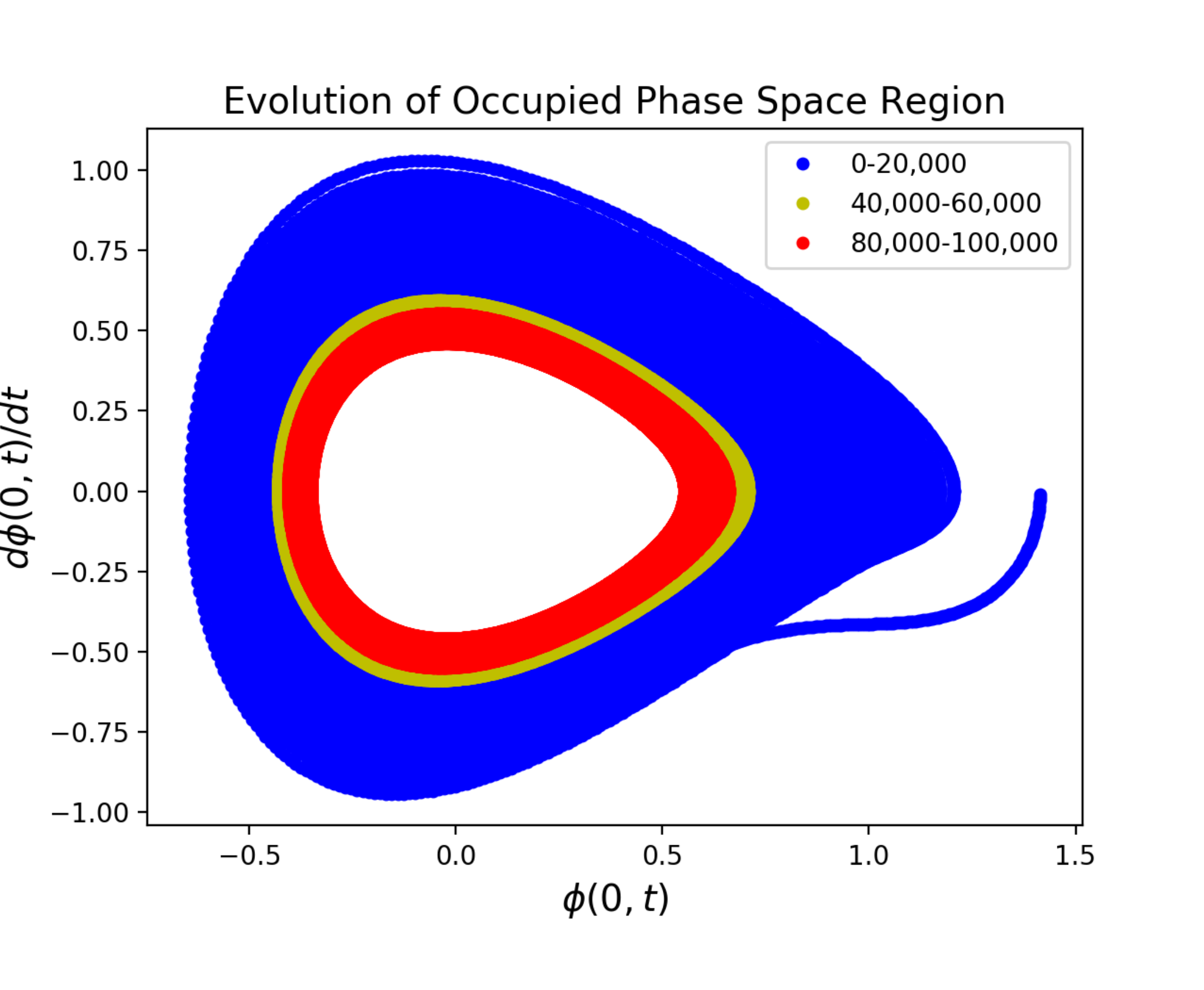}
\caption{Phase space diagram showing that 2d oscillons become more stable as time increases.  This plot is for the oscillon with $r_0 = 3.1$.}
\label{phasediagram2}
\end{figure}

To examine the time evolution of 2d oscillons, in Figure \ref{phasediagram2} we plot the phase space diagram of a single oscillon with $r_0 = 3.1$ in different temporal increments, corresponding to different colors.  From 0 to 20,000 time units, the oscillon moves through the red, yellow, and blue regions of phase space. For intermediate times from 40,000 to 60,000 time units, the oscillon occupies both the red and yellow regions. For longer times between 80,000 and 100,000 time units, the oscillon can only be found in the red region. These results correlate with the increased stability of the oscillon, and were reproduced for all oscillons we investigated.


\section{Summary and Conclusions}

The intriguing possibility that certain spherically-symmetric oscillons in three-dimensional relativistic scalar field theories could be either extremely long-lived or stable has been conjectured a while back by Honda and Choptuik \cite{Honda}. The same possibility was suggested in Ref. \cite{GleiserSorn} for radially-symmetric oscillons in two spatial dimensions. In the present paper, we offer numerical support suggesting  that the conjectures may be correct in both three and two spatial dimensions. We hope that our results will inspire further work on this very intriguing possibility. To arrive at this conclusion, we applied the same two methods in both cases: a parametric resonance analysis and a study of the oscillons' increased virialization for long time scales. 

As shown by HC, three-dimensional oscillons exhibit a rich resonance structure based on their lifetime versus the initial radius $r_0$ of the configuration used as an initial condition to generate them; in this case, Gaussian field profiles of the form $\sim \exp(-r/r_0)^2$. (See Figure \ref{resmountain}.) It is those ``resonant oscillons,'' the ones at the peak of each resonance, that are conjectured not to decay. Although it is impossible to study them numerically directly--it would involve infinite fine-tuning, just as the critical point of a continuous phase transition (as it is, we already must go to eleven decimal places)--we can study the trend toward infinite longevity as we climb each individual resonance towards its peak. 

In 3d, the parametric resonance analysis, summarized in Figure \ref{floquet}, produces three essential results: first, on both sides of a single resonance (super and subcritical oscillons) the Floquet exponents responsible for amplifying the oscillation amplitude of the field configuration have similar qualitative values. This can be seen comparing Figs. \ref{floquet}a and \ref{floquet}b. Second, still on a single resonance, the Floquet exponents increase as lifetimes increase. Third, the Floquet exponents increase as we probe higher resonances along the ``resonance mountain,'' as do the lifetimes related to the resonances' base values. Thus, larger Floquet exponents correlate with longer lifetimes. Given that larger Floquet exponents imply larger amplitude fluctuations for the field, these results indicate that parametric resonance drives oscillations into the nonlinear regime, extending the configuration's lifetime. This idea had been explored in CGM to obtain the minimum radius needed to form an oscillon from an initial Gaussian profile. The trend in the exponents is not exact, however. This can be due to the fact that parametric resonance relies on a linear approximation to the mode equation, and nonlinearities do play a role. Still, the overall trend is clear from the Figure and supports the HC conjecture. We note that previous studies of oscillons under nonspherical classical pertubations showed that these radiated away quickly, confirming the robustness of such solutions \cite{Adib}. It would be interesting to examine the behavior of 3d resonant oscillons under such perturbations, something we leave for future work.

Figure \ref{deltavir} clearly shows that the departure from virialization decreases as the lifetime of an oscillon along a given resonance increases. Requiring very fine resolution, the computation indicates that longer-lived oscillons are better virialized, and hence closer to a harmonic behavior as should be expected.

In 2d, the situation is different given that all oscillons are, in principle, classically stable and there are no special resonances. In a sense, each oscillon is a resonance. To investigate their longevity we used the fact that different oscillons have different plateau energies. A parametric resonance analysis showed that the Floquet exponents for different oscillons grow with plateau energy, indicating that the amplification for their oscillation amplitudes grows with $r_0$. This correlates with previous results from Ref. \cite{GleiserSicilia}, where it was proposed that there is an attractor oscillon in 2d. In our study, this oscillon is well approximated by the one emerging from $r_0=2.42$, which has the lowest plateau energy. (One may think of it as the ground state of a continuous spectrum of oscillons.) Higher initial radii lead to larger initial instabilities that require larger amplitude fluctuations to be neutralized. This simple picture is strongly supported by two results: first, there is a decrease in the value of the Floquet exponent in time, indicating that the oscillon is approaching the attractor (see Figure \ref{floquetlong}). Second, a virial study clearly shows that this is indeed the case, and that the oscillon with $r_0=2.42$ is the better virialized (see Figure \ref{relvir}). To corroborate these results, a phase space study shows that, indeed, oscillons with larger radii will slowly drift into the same phase-space volume as the lowest energy one. 

Finally, we note that parametric resonance and departure from virialization are not the only measures we can use to study the stability and longevity of oscillons.  Configurational entropy, denoted CE \cite{GleiserStamatopoulos1}, is an information measure that has been shown to correlate well with the stability of systems \cite{GleiserSowinski1, GleiserJiang1, Correa, Bernardini, Braga}.  In particular, it was recently shown to be able to predict an oscillon's lifetime with knowledge of its very early dynamical behavior \cite{GleiserStephens}. We intend to study resonant oscillons in 3d and 2d through a CE framework as another test of their longevity and stability.

\acknowledgments
We would like to thank Damian Sowinski for the original implementation of our code and many discussions and suggestions in the early stages of this work, and Michelle Stephens for reading and editing several drafts of this manuscript and for helping to review our calculations. MG and MK are supported in part by a Department of Energy grant DE-SC0010386.


\begin{thebibliography}{99}

\bibitem{Bogolyu} 
Bogolyubskii, I.L. and Makhan'kov, V.G.,
JETP Lett\ {\bf 25}, no. 2, 107-110 (1977)

\bibitem{Gleiser94} 
  M.~Gleiser,
  Phys.\ Rev.\ D {\bf 49}, 2978 (1994)
  doi:10.1103/PhysRevD.49.2978
  [hep-ph/9308279].
  
\bibitem{Copeland} 
  E.~J.~Copeland, M.~Gleiser and H.-R.~Muller,
  Phys.\ Rev.\ D {\bf 52}, 1920 (1995)
  doi:10.1103/PhysRevD.52.1920
  [hep-ph/9503217].
  
\bibitem{Honda} 
  E.~P.~Honda and M.~W.~Choptuik,
  Phys.\ Rev.\ D {\bf 65}, 084037 (2002)
  doi:10.1103/PhysRevD.65.084037
  [hep-ph/0110065].

\bibitem{IkedaYooCardoso} 
  T.~Ikeda, C.~M.~Yoo and V.~Cardoso,
  Phys.\ Rev.\ D {\bf 96}, no. 6, 064047 (2017)
  doi:10.1103/PhysRevD.96.064047
  [arXiv:1708.01344 [gr-qc]].
  
\bibitem{GleiserGrahamStama1} 
  M.~Gleiser, N.~Graham and N.~Stamatopoulos,
  Phys.\ Rev.\ D {\bf 82}, 043517 (2010)
  doi:10.1103/PhysRevD.82.043517
  [arXiv:1004.4658 [astro-ph.CO]].
  
\bibitem{GleiserGrahamStama2} 
  M.~Gleiser, N.~Graham and N.~Stamatopoulos,
  Phys.\ Rev.\ D {\bf 83}, 096010 (2011)
  doi:10.1103/PhysRevD.83.096010
  [arXiv:1103.1911 [hep-th]].

\bibitem{Amin1} 
  M.~A.~Amin, R.~Easther, H.~Finkel, R.~Flauger and M.~P.~Hertzberg,
  Phys.\ Rev.\ Lett.\  {\bf 108}, 241302 (2012)
  doi:10.1103/PhysRevLett.108.241302
  [arXiv:1106.3335 [astro-ph.CO]].
  
\bibitem{Amin2} 
  M.~A.~Amin, R.~Easther and H.~Finkel,
  JCAP {\bf 1012}, 001 (2010)
  doi:10.1088/1475-7516/2010/12/001
  [arXiv:1009.2505 [astro-ph.CO]].

\bibitem{Amin3} 
  M.~A.~Amin and D.~Shirokoff,
  Phys.\ Rev.\ D {\bf 81}, 085045 (2010)
  doi:10.1103/PhysRevD.81.085045
  [arXiv:1002.3380 [astro-ph.CO]].
  
\bibitem{Amin4} 
  K.~D.~Lozanov and M.~A.~Amin,
  Phys.\ Rev.\ D {\bf 97}, no. 2, 023533 (2018)
  doi:10.1103/PhysRevD.97.023533
  [arXiv:1710.06851 [astro-ph.CO]].
  
\bibitem{GleiserSorn} 
  M.~Gleiser and A.~Sornborger,
  Phys.\ Rev.\ E {\bf 62}, 1368 (2000)
  doi:10.1103/PhysRevE.62.1368
  [patt-sol/9909002].

\bibitem{SalmiHindmarsh} 
  P.~Salmi and M.~Hindmarsh,
  Phys.\ Rev.\ D {\bf 85}, 085033 (2012)
  doi:10.1103/PhysRevD.85.085033
  [arXiv:1201.1934 [hep-th]].
  
  \bibitem{Fodor1}
  G. Fodor, P. Forgács, Z. Horváth, and \'A. Luk\'acs,
  Phys. Rev. D {\bf 78}, 025003 (2008).
  [arXiv:0802.3525]].
  
  \bibitem{Fodor2}
  G. Fodor, P. Forgács, Z. Horváth, and M. Mezei
  Phys. Lett. B {\bf 674}, 319-324 (2009).
  [arXiv:0903.0953 [hep-th]].
  
\bibitem{GleiserSicilia}  M. Gleiser and D. Sicilia, Phys. Rev. Lett. {\bf 101}, 011602 (2008).
  
\bibitem{SaffinTranberg} 
  P.~M.~Saffin and A.~Tranberg,
  JHEP {\bf 0701}, 030 (2007)
  doi:10.1088/1126-6708/2007/01/030
  [hep-th/0610191].
  
\bibitem{GleiserHowell} 
  M.~Gleiser and R.~C.~Howell, Phys. Rev. E {\bf 68}, 065203(R) (2003), arXiv:hep-ph/0209176.
  
\bibitem{MathieuBook},
  N.~W.~Mclachlan,
  \enquote{Theory And Application Of Mathieu Functions},
  Dover Publications; Reprint edition (1964)
  
  \bibitem{Adib}
  A. B. Adib, M. Gleiser, and C. A. S. Almeida, Phys. Rev. D {\bf 66} (2002) 085011.
  
 \bibitem{GleiserStamatopoulos1} M. Gleiser and N. Stamatopoulos, Phys. Lett. B {\bf 713}, 304 (2012), arXiv:1111.5597 [hep-th].
  
 \bibitem{GleiserSowinski1} M. Gleiser and D. Sowinski, Phys.\ Lett.\ B {\bf 727}, 272 (2013), arXiv:1307.0530 [hep-th].

\bibitem{GleiserJiang1} M.~Gleiser and N.~Jiang, Phys.\ Rev.\ D {\bf 92}, 044046 (2015), arXiv:1506.05722 [gr-qc].

\bibitem{Correa} R. A. C. Correa, R. da Rocha,  Eur. Phys. J. C {\bf 75}, 522 (2015), arXiv:1502.02283v2 [hep-th]

\bibitem{Bernardini} A. E. Bernardini, N. R. F. Braga, R. da Rocha, Phys. Lett. B {\bf 765}, 81 (2016),  arXiv:1609.01258 [hep-ph].

\bibitem{Braga} N. R. F. Braga, R. da Rocha, Phys. Lett. B {\bf 767}, 381 (2017),  arXiv:1612.03289v2 [hep-ph].

  
\bibitem{GleiserStephens} M. Gleiser, M. Stephens, and D. Sowinski, Phys. Rev. D {\bf 97}, 096007 (2018), arXiv:1803.08550v1 [hep-th].


\end{thebibliography}
\end{document}